\documentclass[fleqn,10pt]{wlscirep}
\usepackage{amsmath,amssymb,amsthm}
\usepackage{graphicx,float,subfig}
\usepackage{caption}
\usepackage{listings}
\usepackage{xcolor}
\usepackage{url}
\usepackage{textcomp}
\usepackage{comment}
\usepackage{multirow}
\usepackage[labelsep=period]{caption}
\DeclareCaptionLabelFormat{adja-page}{\hrulefill\\#1 #2 \emph{(previous page)}}
\usepackage{subfig}
\begin{document}

\title{Analysis of bio-electro-chemical signals from passive sweat-based wearable electro-impedance
spectroscopy (EIS) towards assessing blood glucose modulations } 
\author[1]{Devangsingh Sankhala}
\author[2]{Madhavi Pali}
\author[2]{Kai-Chun Lin}
\author[2]{Badrinath Jagannath}
\author[3]{Sriram Muthukumar*} 
\author[2]{Shalini Prasad*} 

\affil[1]{Department of Electrical Engineering, The University of Texas at Dallas, 800 W Campbell Rd, Dallas 75080 TX, USA}
\affil[2]{Department of Bioengineering, The University of Texas at Dallas, 800 W Campbell Rd, Dallas 75080 TX, USA}

\affil[3]{EnLiSense LLC, 1813 Audubon Pond Way, Allen 75013 TX, USA}




\begin{abstract}
There has been a recent tremendous interest in label-free detection of biomarkers which is a critical enabler of point-of-need diagnostics.A low-power, small form factor, multiplexed wearable system is proposed for continuous detection of glucose in passively expressed sweat using electrochemical impedance spectroscopy (EIS) measurement. The wearable EIS system consists of a sensing analog front end integrated with a low-volume (1-5 $\mathrm{\mu}$L) ultra-sensitive flexible biosensors. A passive sweat sensor was designed integrating  glucose oxidase electrochemical system on an active semiconducting material. The non-faradaic EIS response of the biosensor was used to calibrate the analog front end's response using ratiometric Discrete Fourier Transform (DFT) for a shorter measurement time. In this work, a stringent assessment of a continuous glucose sensing platform is performed in a bottom-up approach, going from the biosensor to the system system to the interaction with a human subject. The active semiconductor-based biosensors are dosed with glucose concentrations ranging from 5-200 mg/dL and detection is performed using the analog front end. In addition, detailed analysis of battery life and performance of a wearable EIS system is discussed to define a figure of merit for an optimally integrated design. Moreover, a continuous glucose detection test is performed on a healthy human subject cohort to investigate the stability of sensor-system mechanism for an 8-hour period and a time-series based, auto-regressive (AR) model was created for the system.
\end{abstract}

\flushbottom
\maketitle
\thispagestyle{empty}


\section*{Introduction}
There is a significant interest in wearable biosensing platforms in the recent times. Constant improvements are being made in developing accurate and robust sensing mechanisms for various biomolecules of interest that demonstrate temporal dynamics. Wearable biosensors are enablers for point of need tracking of chronic conditions and can be beneficial both to the user as well as the healthcare professional in guided therapy~\cite{wang2001glucose}. Glucose is a key biochemical marker with temporal dynamics in context to monitoring chronic health condition of diabetic patients. Current technologies to monitor blood glucose in small and/or wearable form factors still require frequent finger pricks or needle based interstitial blood glucose monitoring methods such as the Abbott Freestyle Libre or the Dexcom G6 to be integrated with pedometers and other lifestyle management tools towards achieving patient actionable glycemic control. Glucose biosensors, found commercially as well as in literature, implement a microneedle-based, chronoamperometry~\cite{lee2016graphene,bhansali,kimAlcohol,gaoMUX} and square-wave voltammetry~\cite{davidSWRalcohol} biosensor compatible for wearable devices, due to ease of electrical circuit implementation. These sensors may provide good sensitivity at the cost of larger sample volumes of interstitial fluid or sweat to adhere to acceptable CGM standards~\cite{cgmstdBattelino2019clinical}. Moreover, due to larger settling time of current, density gradients cause undesirable convective disruption of the diffusion layer~\cite{bardFaukner}. In addition, large settling time of an amperometric sensing scheme results in relatively higher current consumption.

 There is an immense opportunity in designing a wearable electronic system towards supporting such point-of-need sensing applications with low-volume, true no-needle insertion mechanisms. A case study of such a system is performed here by coupling a low-volume passive sweat biosensor~\cite{rm_glucose} with a rapid response electrochemical impedance spectroscopy (EIS)-based system~\cite{SankhalaCortisol}. The proposed glucose reporting mechanism utilizes passively expressed system sweat towards achieving quantitative decision making; therefore low sample volume makes it prone to erroneous measurement in a real-world scenario. Error can be defined as a property of the system which introduces a deterministic undesirable offset in the measured outcome of the device. In this context, deterministic offset would mean an offset that can be mathematically represented using a polynomial. On the other hand, noise is a probabilistic property of the system which introduces a random offset in the measured outcome of the system and is modeled as a probability distribution function. The presence of relatively higher noise and error affects the battery performance of a device as it would require more samples to average or filter out the undesirable signal counterpart. In this work, the sensor's impedance change is reported and correlated to the conventional statistical methods to establish good sensitivity. Continuous sensing of glucose dose concentration and differentiation of normal and hyperglycemic rate of change of glucose is demonstrated in a healthy cohort of 20 subjects, using an auto-regressive model, in comparison to an interpolated sweat glucose time series.

\section*{Results}
\subsection*{Calibration of the system using lumped elements }
\begin{figure}
\centering
	\subfloat[]{{\includegraphics[width=0.42\textwidth]{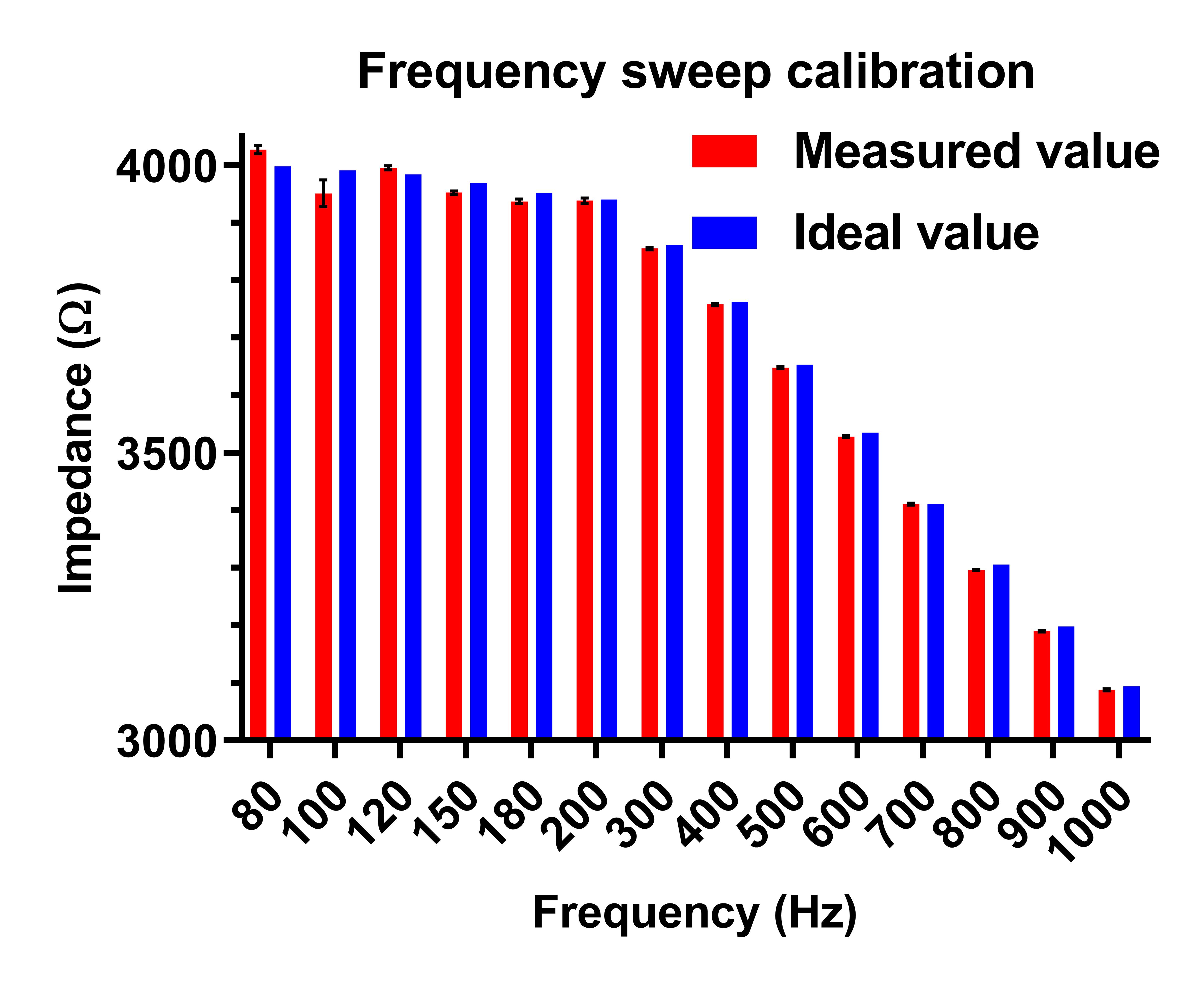} }\label{sweep}}
    	\quad
    \subfloat[]{{\includegraphics[width=0.42\textwidth]{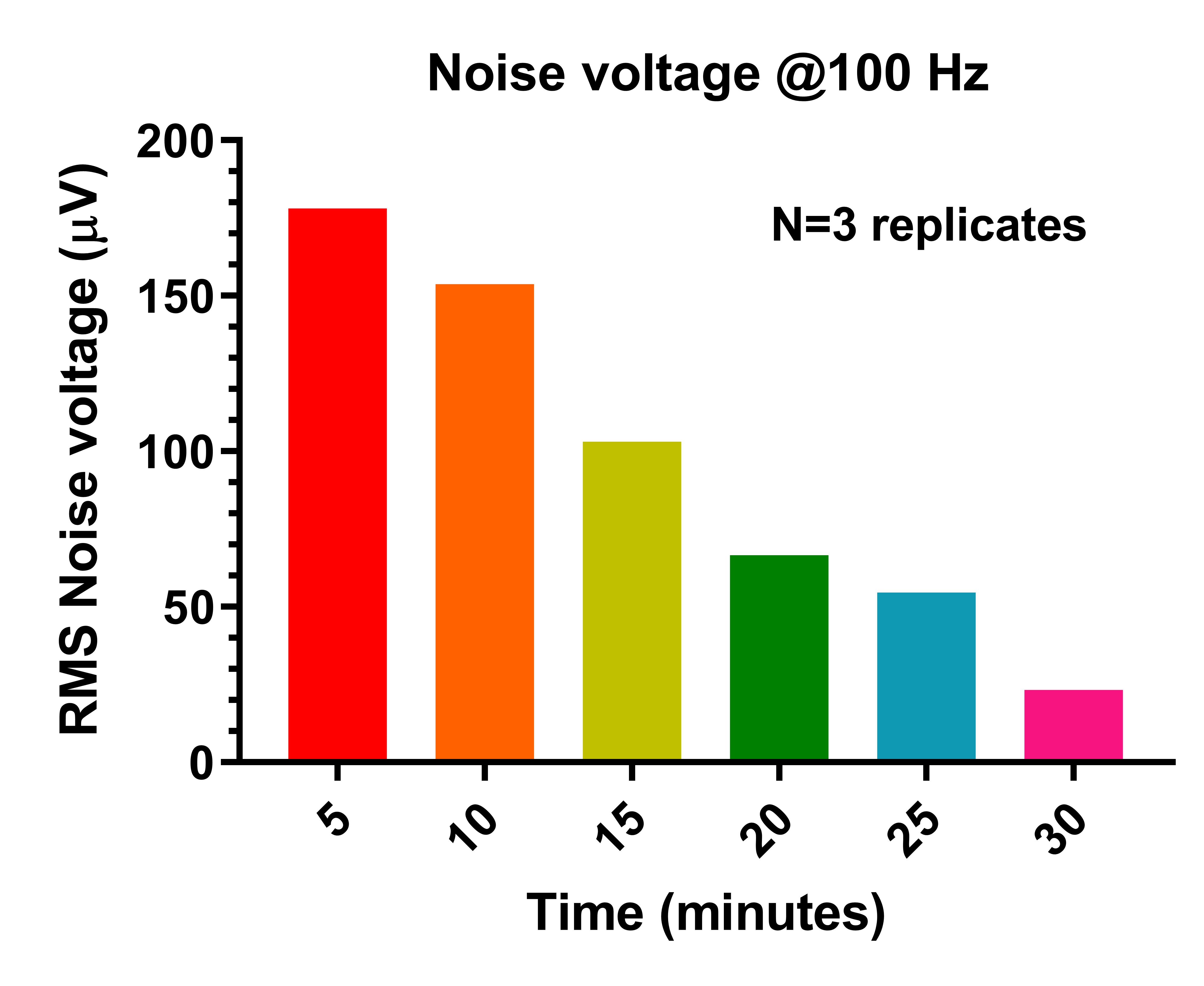} }\label{noiseAt100}}
   \quad        
    \subfloat[]{\includegraphics[width=.5\textwidth]{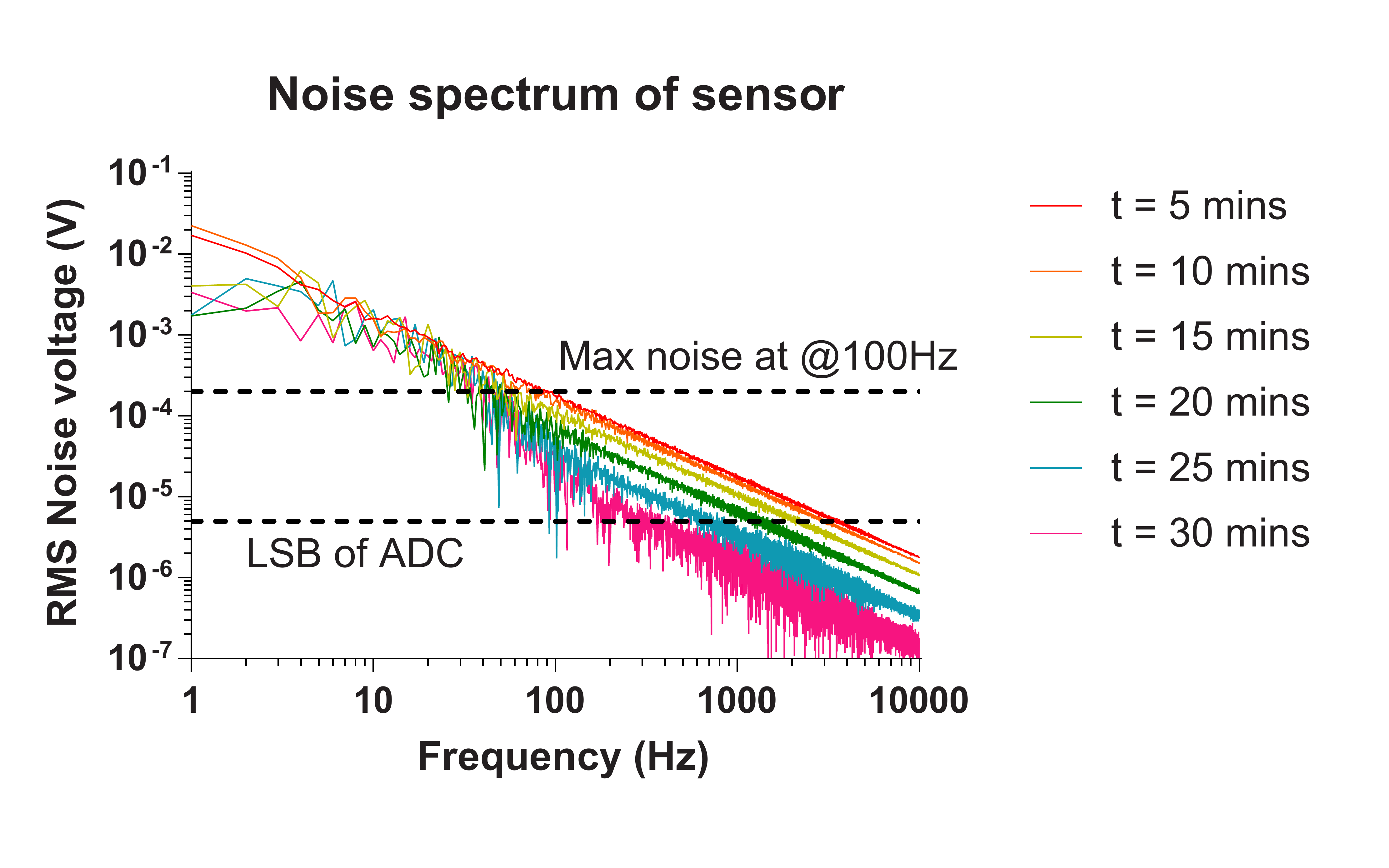}\label{noiseSpectra}} 
      \quad
      \subfloat[]{{\includegraphics[width=0.45\textwidth]{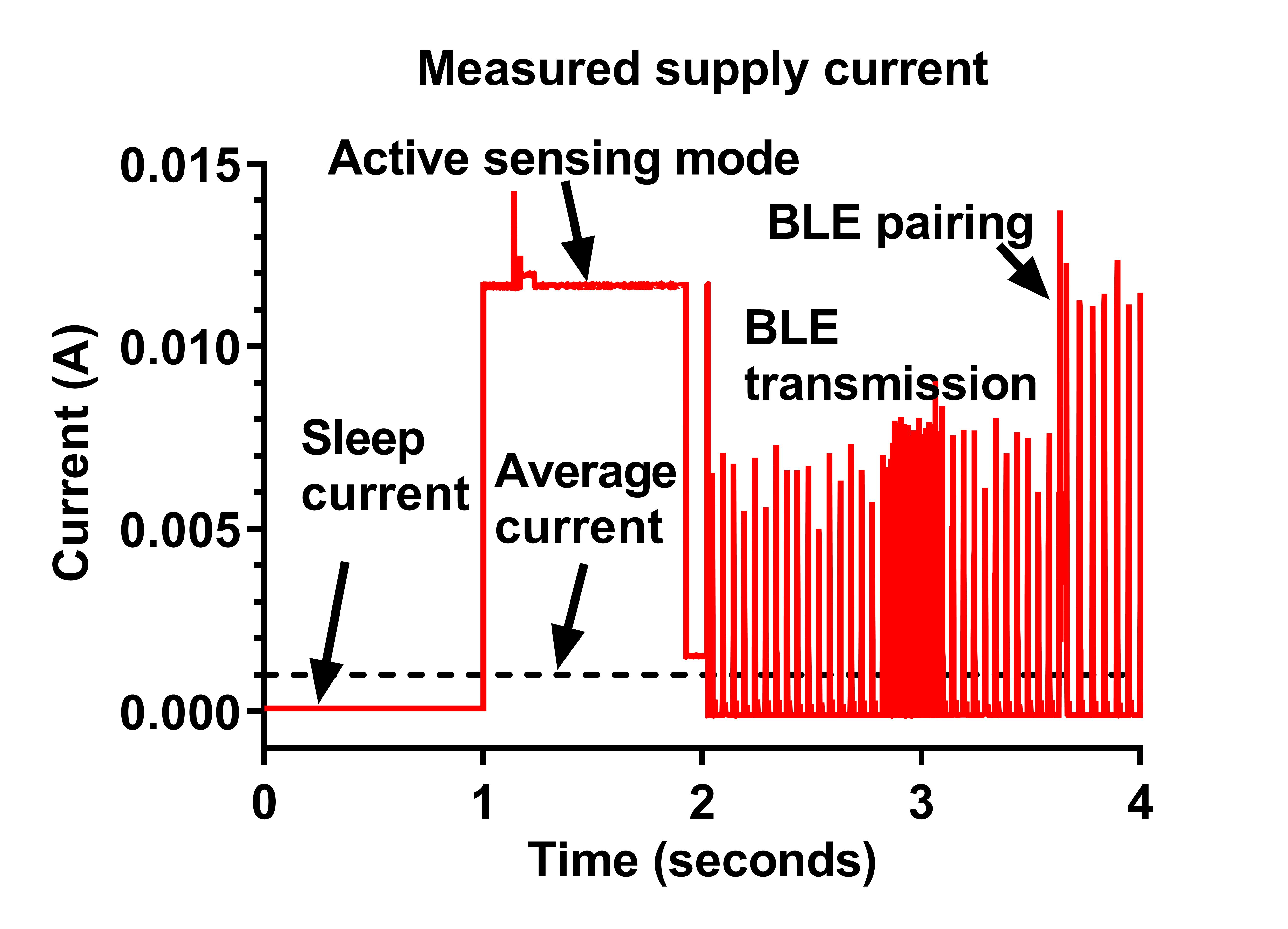} }\label{current}}        
    \caption{Impedance of a universal test cell was measured using n = 3 replicates of the wearable EIS device.~\ref{sweep} Spread of measured impedance across all device replicates at frequencies ranging from 80 Hz to 1 kHz demonstrate reasonable accuracy with respect to the PSPICE simulated impedance. ~\ref{current} Current consumption was measured for the proposed EIS device using the Agilent N6705B power analyzer. Various operating modes of the device are shown with their characteristic current consumptions.~\ref{noiseSpectra} Noise spectrum of a biosensor measured using a high speed potentiostat while introducing PBS on the non-functionalized sensor surface.~\ref{noiseAt100} Measured RMS noise amplitude of the sensor surface at 100 Hz shows a decrease in cumulative sensor noise with time.}\label{calibData}
\end{figure}

The proposed system was tested for individual channel offset on the calibration channel of a universal dummy cell (Gamry Instruments). Fig.~\ref{calibData} shows the variation of measured impedance of the universal dummy cell using n = 3 replicates of that same system. It is seen that the measured impedance. The accuracy of the device is demonstrated in Fig.\ref{sweep} over n = 3 devices over a single channel. The expected calibration impedance at $100 Hz$ was $4 k\Omega$. However, a systematic negative offset of $50\Omega$ is observed with a variability of $50\Omega$ on the measured impedance. This accounts for a $1.25\%$ inter-device variation along with a $-1.25\%$ error. The precision of a single device replicate is observed by taking n = 100 measurements for all four channels using the universal dummy cell. The peak inter-channel variability of $~150\Omega$ is observed.

\subsection*{Calibration dose response of the system}
\begin{figure}
\centering
	\subfloat[]{{\includegraphics[width=0.42\textwidth]{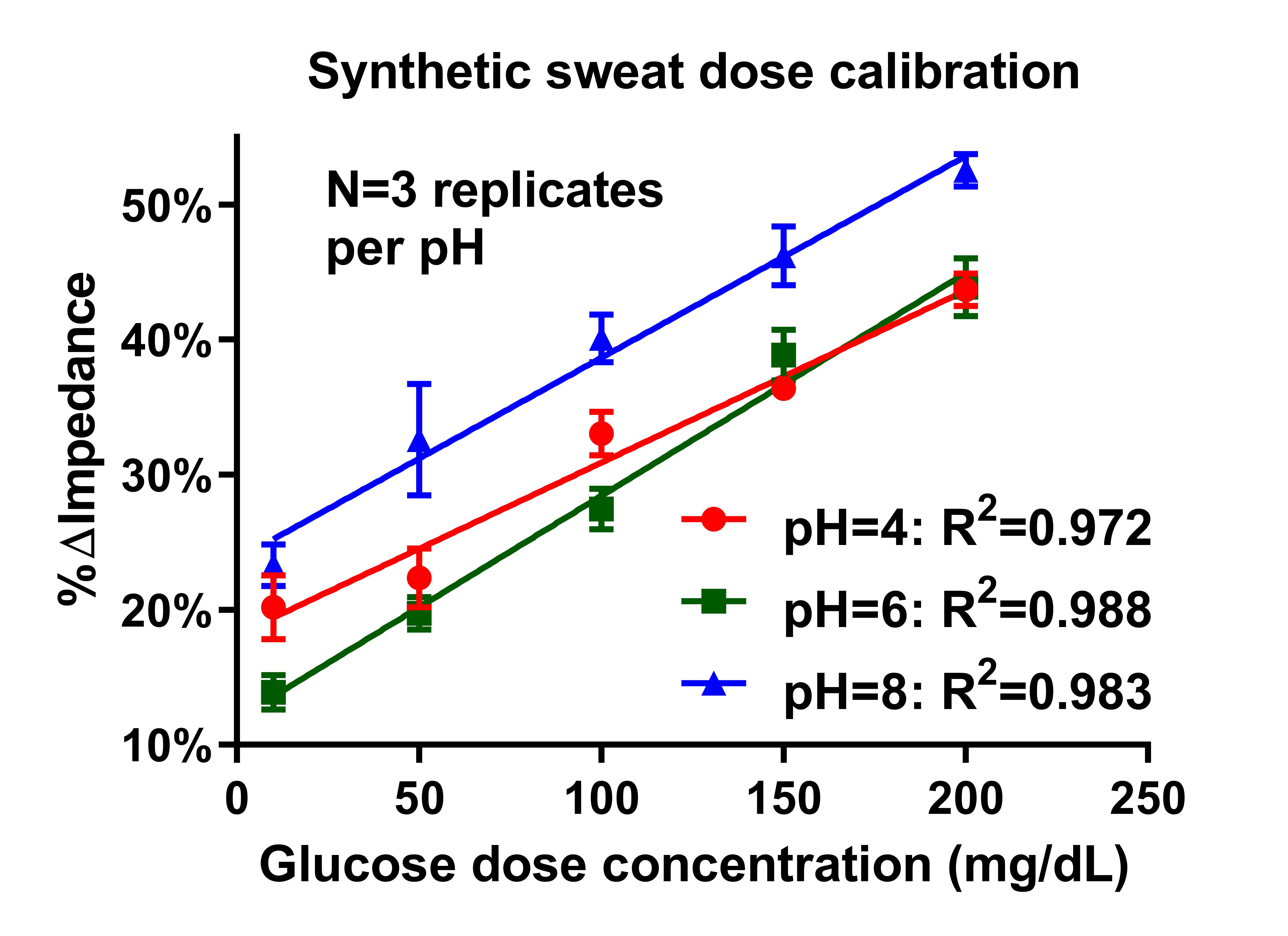} }\label{CDR}}
    \quad
    \subfloat[]{{\includegraphics[width=0.42\textwidth]{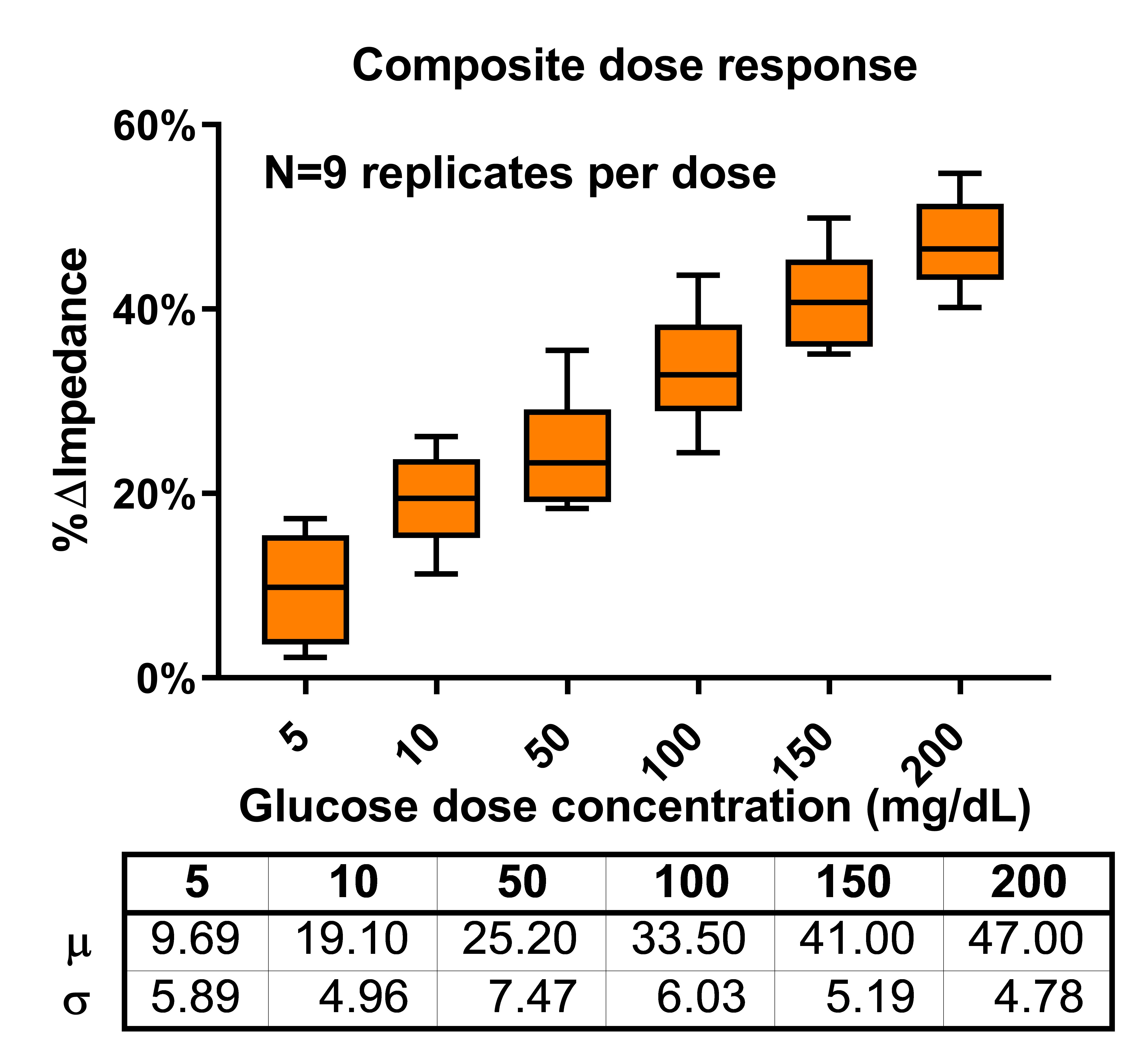} }\label{boxPlot}}
    \caption{~\ref{CDR} Calibration dose response of the system using synthetic sweat buffer of various pH.~\ref{boxPlot} A box-whisker plot is used to observe the true variability of a given dose concentration as a composite of all buffers used in the work.} \label{CDRplots}
\end{figure}

Glucose dose concentrations ranging from 5-200 mg/dL were made in synthetic sweat buffer with pH values 4, 6 and 8 were introduced on a functionalized sensor and the impedance at $100Hz$ was recorded to demonstrate proof of binding interaction on the sensor surface. Fig.~\ref{CDR} shows the response of the system as standard error of mean of the percentage change function of impedance impedance to varying pH and dose concentrations with a linear fit. Eqn.~\ref{deltaZ} shows the mathematical percentage change of the measured impedance used to plot dose response. The system response shows $~3\%$ standard error on most of the dose concentrations thus demonstrating precision of measurement. 
\begin{equation}\label{deltaZ}
\text{Percentage change in Z} = \%\Delta Z= \dfrac{Z - Z_{baseline}}{Z_{baseline}}
\end{equation}

Box-whisker plots were created as shown in Fig~\ref{boxPlot} to investigate the variability of the system response for a given dose concentration irrespective of buffer pH. The true variability of the sensor response was observed to be $~10\%$ inter-sample variation in the form of inter-quartile region height of each of the glucose dose concentrations. It can be observed that even with a larger variability, the median divides the inter-quartile region symmetrically, demonstrating a Gaussian distribution. However, there is a need to model this variation as a mathematical relationship to allow accurate detection as well as differentiation of glucose dose concentrations across the dynamic range.

\subsection*{Identification of sources of noise and error}
Noise is an undesirable measured quantity that accompanies a true measured quantity. This introduces an amount of uncertainty in the accuracy and repeatability of the measurement device. Sources of noise have been well characterized for CMOS devices in the past~\cite{Hooge,thermalNoise}.  Primarily, the two major sources of noise discussed in academic as well as industrial design are thermal noise and flicker noise for  complementary metal-oxide-semiconductor (CMOS) transistors. Thermal noise is a result of electrical fluctuations caused by thermal energy~\cite{thermalNoise}. This electron motion is popularly known as Brownian motion. The RMS noise voltage is expressed as in Eqn.~\ref{thermalNoiseEqn} where $k$ is the Boltzmann constant, $T$ is the absolute temperature in kelvin and $R$ is the resistance of the device under test. Thus, thermal noise will increase with rise in temperature. In CMOS transistors, this noise is a summation of the noise generated by the gate and channel resistance. In case a capacitor is connected in parallel with the resistive noise source, the RMS thermal noise is denoted as in Eqn.~\ref{thermalNoiseRC}. Thus, thermal noise can be controlled using a shunt capacitor~\cite{razavi}.
\begin{equation}\label{thermalNoiseEqn} 
\bar{v} _{n,thermal}= \sqrt{4kTR }
\end{equation}
\begin{equation}\label{thermalNoiseRC}
\bar{v} _{n,thermal,RC}=\sqrt[]{\dfrac{kT}{C} }
\end{equation}
 Flicker noise or popularly known as 1/f noise is the resultant of attractive forces at the silicon-oxide interface in a CMOS transistor giving rise to several energy states~\cite{razavi,Hooge}. The RMS flicker noise is expressed as in Eqn.~\ref{fNoiseEqn} where $K$ is a statistically calculated, process dependent parameter, $W$ and $L$ are the dimensions of the transistor, $C_{ox}$ is the capacitance of oxide film per square and $f$ is the frequency of interest. Thus, 1/f noise is a function of frequency, and is a source of error for low frequency circuits.
\begin{equation}\label{fNoiseEqn}
\bar{v}_{n,1/f} = \sqrt[]{\dfrac{K}{WLC_{ox}}.\dfrac{1}{f}}
\end{equation}

\begin{figure}
\centering	
      \subfloat[]{{\includegraphics[page=1,viewport= 365 710 530 830,clip=true,scale=.6]{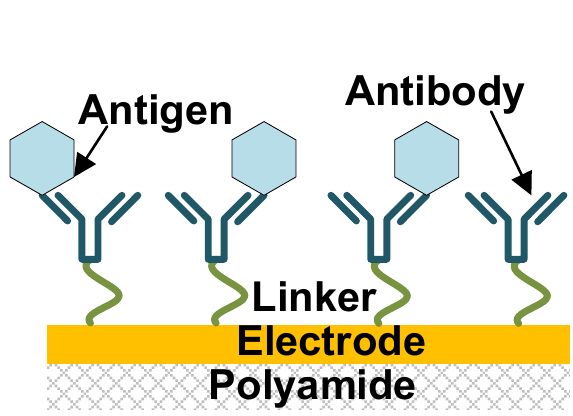} }\label{molecular}}
      \quad
      \subfloat[]{{\includegraphics[page=1,viewport= 40 390 170 670,clip=true,scale=0.6]{imagepdf} }\label{sensorCS}}
      \quad      
      \subfloat[]{{\includegraphics[page=1,viewport= 190 390 320 700,clip=true,scale=.7]{imagepdf} }\label{sensorCkt}}
      \quad
      \subfloat[]{{\includegraphics[page=1,viewport= 330 390 550 700,clip=true,scale=.7]{imagepdf} }\label{sensorCktnoise}}
      \quad
    \subfloat[]{{\includegraphics[page=2,viewport= 50 530 570 780,clip=true,scale=0.52]{imagepdf} }\label{blockdia}}  
      \subfloat[]{{\includegraphics[width=0.3\linewidth,angle=90]{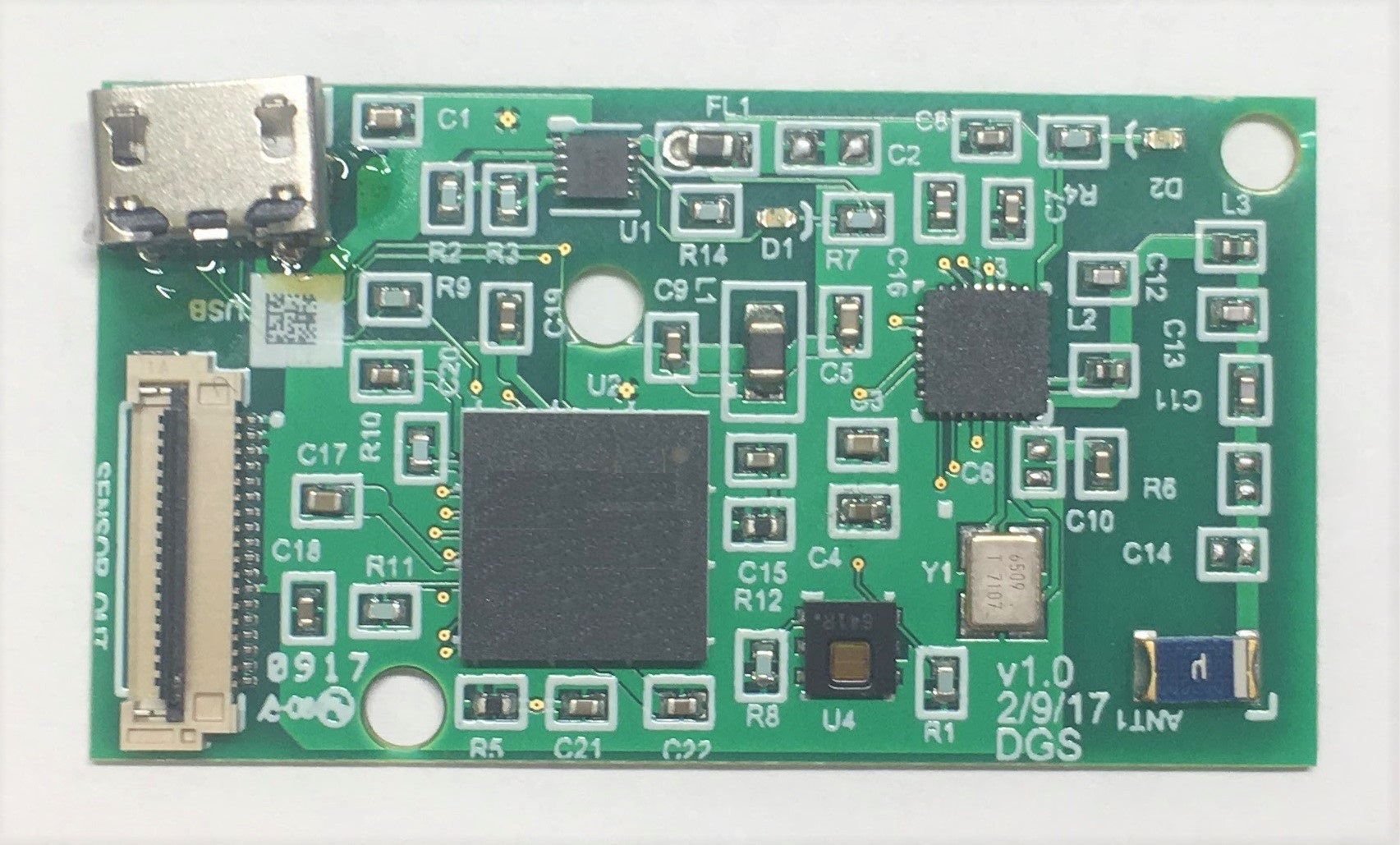} }\label{mbx}}
    \quad
    \subfloat[]{{\includegraphics[scale=0.12]{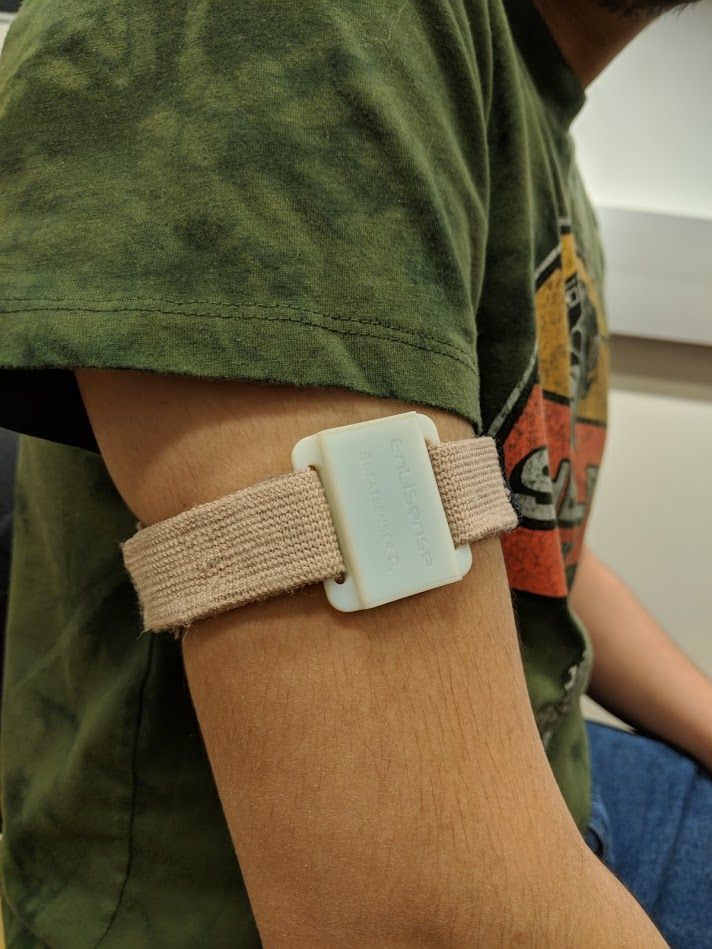} }\label{hand}}
    \caption{  ~\ref{molecular}  Detection mechanism of the glucose immunoassay on the sensor surface is created using glucose oxidase antibody-antigen interaction in the presence of glucose oxidase enzyme. ~\ref{sensorCS}  Cross section of the sensor surface. ~\ref{sensorCkt}  Equivalent incremental circuit modeling arising from sensing mechanism. ~\ref{sensorCktnoise} Equivalent circuit model with respective sources of noise added to the sensing mechanism. Disassembly of the wearable EIS device.  ~\ref{blockdia} Block diagram of the proposed EIS device. ~\ref{mbx} The motherboard used for EIS experiments is equipped with an ADuCM350 SoC and nRF8001 Bluetooth Low Energy (BLE) chip-set. The slide-in connector on the bottom allows for connecting sensors. ~\ref{hand} Wearable EIS device worn as a band.}\label{assay}
\end{figure}

 A noise model proposed by Hassibi and others~\cite{hassibi_noise} is used to provide a noise treatment for the biosensors mentioned in this work. For a non-faradaic electrode-electrolyte interface, the valence of the electrolyte is $z$. In addition, the measurement is performed for a single frequency, thus the bandwidth of the system is 1 Hz. Fig~\ref{sensorCkt} shows the incremental circuit model established using impedance measurements in Munje et. al.~\cite{rm_glucose} and this work. Three noise sources due to $R_s$, $R_{ASA}$ and $R_{ct}$ can be identified using previously mentioned noise equations. Moreover, the measurement device's response time of 13 ms is smaller than the biosensor's response time of 5 minutes. Based on these assumptions~\cite{hassibi_noise}, a noise model was proposed for the biosensor used in this work as shown in Fig.~\ref{sensorCktnoise}, which was derived from Fig.~\ref{sensorCkt}~\cite{rm_glucose} by dividing the cross-section of the sensor in three parts.

 Since an active semiconducting film is a partially conductive film with some resistance based on its doping levels, it is prone to a noise voltage similar to a resistor and can be modeled as a thermal noise source. However, considering semi-conducting properties of the film in mind~\cite{znonoiseBarhoumi2014correlations}, it can be assumed that there is a presence of $1/f$ noise due to this film, which is a result of charge carrier generation as well as recombination. However, since a capacitance $C_{ASA}$ exists in parallel with the noise sources, this noise can be modeled as a $kT/C$ noise of the active semiconducting film. Considering the biosensing electrical double layer and charge transfer resistance, these are modeled to be circuit components for an ideal polarized electrode (IPE), however, they are imperfect in a real scenario. Thus, the noise of this region can be assumed to be a variation in the current density of the charge carriers depending upon the strength of the electrical potential, which in turn depends on the location of the slip plane. However, since the ion motion is slower than an electron's motion, a function $M(\omega)$ is used to model the frequency dependent effect of this current noise source. The solution resistance is a bulk effect and can be assumed to be a resistor, thus the thermal noise effect. The final expression of the noise voltage would be a summation of all these noise sources as given in Eqn.~\ref{noiseEqn}.
\begin{equation}\label{noiseEqn}
\bar{v}^2 = {4kT(R_s+R_{electrode}) + 2zqI.M(\omega).(R_{ct}||\dfrac{1}{j\omega C_{dl}}) + \dfrac{kT}{C_{ASA}}+\dfrac{K}{WLC_{ox}}.\dfrac{1}{f}}
\end{equation}

 The veracity of this model is confirmed using a noise test for 30 minutes on a non-functionalized sensor surface wet with 1X phosphate saline buffer(PBS) using a high-speed potentiostat (Zurich Instruments, Switzerland). Fig.~\ref{noiseSpectra} shows the variation of noise spectra at an interval of 5 minutes. The noise voltage follows a trend inversely proportional the the frequency and is maximum at t=5 minutes, whereas it decreases with time and is at a minimum at t=30 minutes at 100 Hz excitation frequency. Thus, as the introduction of PBS reaches an equilibrium, the noise amplitude of the system start decreasing with respect to 100 Hz excitation frequency. The RMS noise voltage is shown in Fig.~\ref{noiseAt100}. 

 Mean absolute relative difference (MARD)~\cite{NoujaimMARD} was used to emphasizes that noisy measurements are a function of the true measurement and a summation of systematic bias and coefficient of variation, to establish reliability of the wearable biosensor. This expression is given as in Eqn.~\ref{NoisyEqn} where $\hat{G}$ is the noisy measurement, $G$ is the true measurement, $\%b$ is the systematic bias obtained from the measurement system's properties, $\%cv$ is the coefficient of variation among each sample for the given dose step and $\epsilon_n$ is a standard random variable. Thus, the systematic bias will be a constant offset correction where as the standard variable which accounts for variation from external environmental effects in a band limited by $\%cv$. 

\begin{equation}\label{NoisyEqn}
\hat{G}= G (1+\%b+\%cv\times\epsilon_n) 
\end{equation}

 This approach can be used to characterize sources of variability in system performance. Systematic offset can be observed as a result of device accuracy and precision, offset of the measurement device and specific signal threshold of the biosensor. Coefficient of variation arises from sample-to-sample variation and is calculated as a ratio of standard deviation to mean for a given dose concentration. In addition, environmental factors such as temperature, relative humidity, variation in volume dispensed, etc. play an important part in the system's response. However, it would be an extensive effort to characterize for each of them separately and superimpose. A standard normal variable can be used to account for all these variations. Thus, variation is accounted for statistically, whereas systematic offsets are accounted for by observation and nullified by software when using Eqn.~\ref{NoisyEqn}.

\begin{table}[ht]
\centering
\hyphenation{dependent}
\begin{tabular}{|c|p{2.3in}|p{0.9in}|p{2.5in}|}
\hline
\textbf{Variable} & \textbf{Source of error} & \textbf{Value} & \textbf{Comments}\\
\hline
\multirow{2}{*}{\%b} & Non-functionalized biosensor electrical noise ($thermal + 1/f$) & 0.35\% to 2\% & Variation of RMS noise of the sensor with respect to excitation voltage = 10 mV RMS \\
\cline{2-4}
 & Instrumentation loop offset & -1.25\% & Offset in impedance mesaurement for $Z_{cal}=3990\Omega$ at $f=100Hz$  \\
\cline{2-4}
 & Biosensor specific signal threshold & 11\% & Y-intercept=11\% for sensor output impedance response from Fig.~\ref{CDR} at buffer pH = 4.  \\
\hline
\multirow{4}{*}{\%cv} & Biosensor-dependent sample-to-sample variation & 8\% to 10\%& Variation @ $f_s = 100Hz$ observed in Fig.~\ref{boxPlot}. \\
\cline{2-4}
& Instrumentation loop variability& -0.4\% to -1.5\% & Sample-to-sample variation of impedance mesaurement for $Z_{cal}=3990\Omega$ at $f=100Hz$ \\
\hline
\end{tabular}
\caption{\label{tab:errorSummary}A simplified MARD model for the proposed wearable platform was created using sources of error as a percentage component of the device under test in a real world environment in the presence of noise.}
\end{table}

\subsection*{Performance metrics for wearable devices}
There is a need to encompass all important specifications as one numerical expression to represent the performance of an electrical design in question relative to other alternative designs. This numerical expression is called the figure of merit (FOM). Such a design comparison is very popular among data conversion designs~\cite{WaldenADC} and is defined as the energy required for one data conversion. For deciding an expression for the figure of merit of a design, the system was split into the measurement device and the biosensor and their inherent properties were observed.

 For the measurement device, the ultimate objective to be achieved would be very similar to a data conversion system, i.e. achieving a low power consumption per every measurement performed. Similarly, the capability of the device to measure with minimal power will be observed by varying operating conditions. A battery operated device is generally operated at a very low, sleep-mode supply current. However, the device requires a comparatively larger amount of current to operate in an active state. Thus, for such an uneven distribution of current, an average over the time taken by the device to execute the program from start to end. Thus, the battery selection for a fixed average current is performed as follows. Let the sleep current of the device in question is $I_{sleep}$ for time $t_{sleep}$ and the active current of the device is $I_{active}$ for time $t_{active}$. Since the device completes one program cycle, the total time for one program cycle $t_{total}$ is the summation of $t_{sleep}$ and $t_{active}$. 

\begin{equation}
t_{total} = t_{sleep} + t_{active}
\end{equation}
\begin{equation}
I_{avg} = I_{sleep} \times \dfrac{t_{sleep}}{t_{total}} + I_{active} \times\dfrac{t_{active}}{t_{total}}
\end{equation}

The figure of merit (FOM) of an electrical device can be defined as the energy spent to convert a physical sensor signal at an instant of time. However, to demonstrate the increased throughput capacity of the device against power consumption, parameters $n$ and $p$ were included, where $n$ as number of sensors channels available by design and $p$ is the number of points per one measurement cycle. Here, one measurement cycle implies measuring all points until settling for amperometric designs, or one sweep from lowest to highest frequency in impedance spectroscopy designs. For the proposed design, the figure of merit is calculated for $n=4$, $p =10$, $NBITS=16$, $f_s=\dfrac{1}{13ms}$ as one conversion period. The voltage is assumed to be $3.7V$ as of a standard lithium polymer battery. Thus, an FOM of $19 nJ/point$, or $190nJ / sweep$ was obtained.
\begin{equation}\label{FOMdevice}
\text{FOM}_\text{device} = \dfrac{npV_{batt} I_{avg}}{f_{s}\times 2^{NBITS}} 
\end{equation}

 A biosensor is expected to demonstrate a large dynamic range with a low order curve fit equation for lesser computation requirement and low power consumption. Moreover, it is expected that the biosensor is sensitive to the very low concentrations of biomarkers of interest. Thus, the figure of merit (FOM) of a biosensor can be defined as in Eqn.~\ref{FOMsensor}, where a relative percentage value of the specific signal threshold and linear dynamic range are considered. thus, higher the FOM, higher the ease of integration irrespective of dissimilar sensor, measurement device or electrochemical technique used or any combination of the former.
\begin{equation}\label{FOMsensor}
\text{FOM}_\text{sensor} = \dfrac{\%LDR}{\%SST} 
\end{equation}

 The above mentioned figures of merit of individual system sub-blocks define the efficiency of the device to perform under trade-offs under the designer's control. However, if these are inconsistent between two sub-blocks of a system, the integrated system may function inaccurately due to a cumulative effect of errors generated by individual sub-blocks. To ensure maximum integrability of the measurement device and the biosensor, an attempt was made to mathematically define a statistical relationship between sub-block accuracy. Accuracy $A$ of a quantity is defined as the difference between the actual value $x$ of the
quantity and the ideal or true value $X$ of the quantity as shown in Eqn.~\ref{accuracyDefn}. Similarly, mismatch between two quantities $x_1$ and $x_2$ is the difference between the actual ratio of the quantities and the desired ratio of the two quantities as shown in Eqn.~\ref{mismatchDefn}. For the proposed biosensor, the FOM of 11.87 if obtained for $\%LDR=95\%$ and $\%SST=8\%$. 
\begin{equation}\label{accuracyDefn}
A = \dfrac{x-X}{X} = \dfrac{\Delta x}{X}
\end{equation}
\begin{equation}\label{mismatchDefn}
\delta = \dfrac{\dfrac{x_2}{x_1}-\dfrac{X_2}{X_1}}{\dfrac{X_2}{X_1}}
\end{equation}
Both the accuracy and mismatch numbers should be zero for a perfectly non-erroneous measurement, which is a highly unlikely in the presence of sources of variation and noise. Thus, this specification helps a system integrator to analyze the quality of sub-blocks available and investigate methods of error compensation, either by hardware or software. For the proposed system system, for simplicity it is assumed that the since the measurement device and the biosensor are made of different materials, the offsets are uncorrelated, thus additive. The accuracy of the measurement device is 1.25\% as per Fig.~\ref{sweep} whereas the accuracy of the biosensor is $10\%$ as per the Fig.~\ref{boxPlot}. Thus, the mutual mismatch as a resultant of coupling both would be $\delta=-0.875\%$ as per Eqn.~\ref{mismatchDefn}. For an optimal system level design, $\delta\rightarrow 0$ thus demonstrating the robustness of integration.

\subsection*{Continuous measurement testing}
The dose calibration response of the sensor describes the accuracy of the system at a given steady state. In contrast, real-world applications demand the need for providing a steady sensor output irrespective of change in environmental conditions. A method of conversion of impedance measurements to real-time glucose measurements would bridge the gap between the sensor output and the much-needed user information to be generated. Such a multi-variable problem can be solved using predictive methods for time-series based analysis. Various models such as auto-regression\cite{leal2010real,reifman2007predictive} and neural network based approach\cite{dash2020neurovad,mhaskar2017deep,naumova2012meta,sivananthan2011assessment,pappada2011neural} have been used, as they consider the previous input, the previous output and the current input of the system. Also, these models have been created using interstitial fluid (ISF) based continues glucose monitoring systems (CGM), hence these are not based on non-invasively sampled data sets.

A more simplistic approach is the application of time-series analysis to measurements provided by the proposed system. The system was put on N=20 human subjects for a period of 8 hours, wherein their glucose sensor output, skin temperature, skin humidity was measured continuously while 4 sweat samples were collected per subject. The sweat sample was used to measure sweat glucose levels using a reference method. Since only 4 reference points were available considering the limitations of the IRB, an interpolation curve was created using the 4 reference points to match the length of output of the system. 

\begin{equation}\label{ARIMAeqn1}
y = X_1\beta_1 + ... +  X_{10}\beta_{10} +\mu_t
\end{equation}
\begin{equation}\label{ARIMAeqn2}
(1-\phi_1 L - ... -\phi_{10} L^{10})\mu_t = (1+\theta_1L + \theta_2L^2 + \theta_3L^3)\epsilon_t
\end{equation}

\begin{figure}
\centering
	\subfloat[]{{\includegraphics[width=0.42\textwidth]{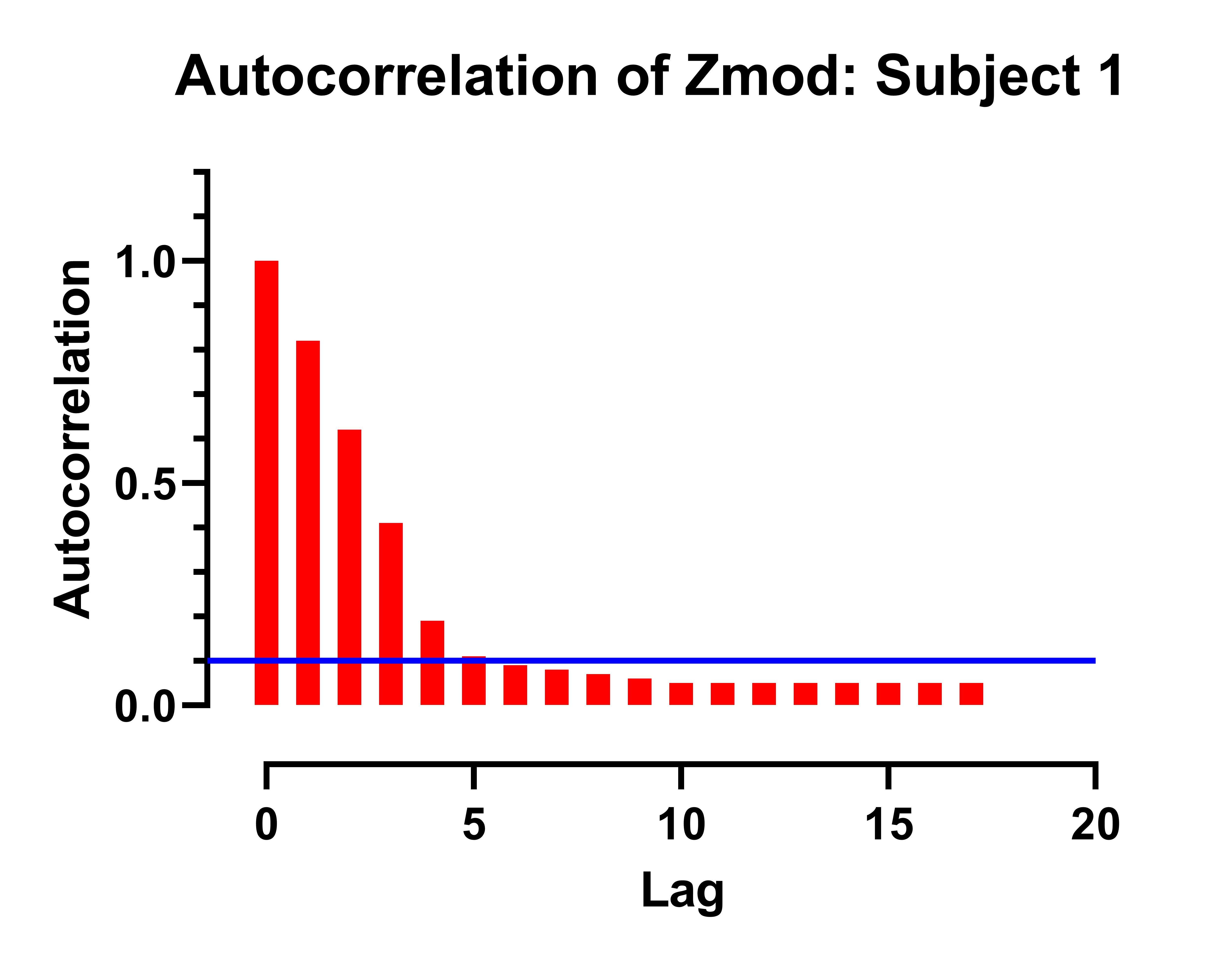} }\label{acfZ1}}
    \quad
    \subfloat[]{{\includegraphics[width=0.42\textwidth]{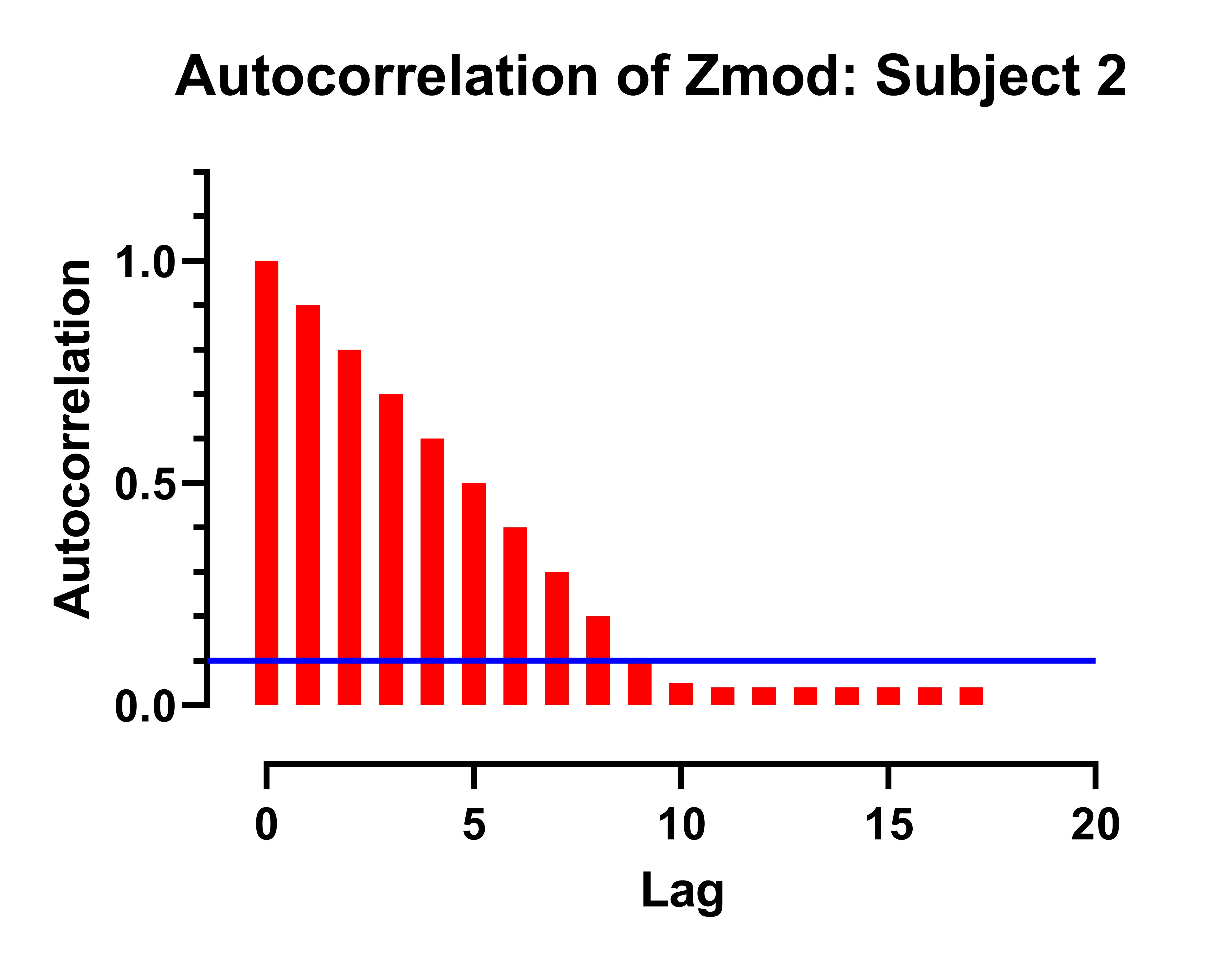} }\label{acfZ2}}
    \quad
    
    \subfloat[]{{\includegraphics[width=0.42\textwidth]{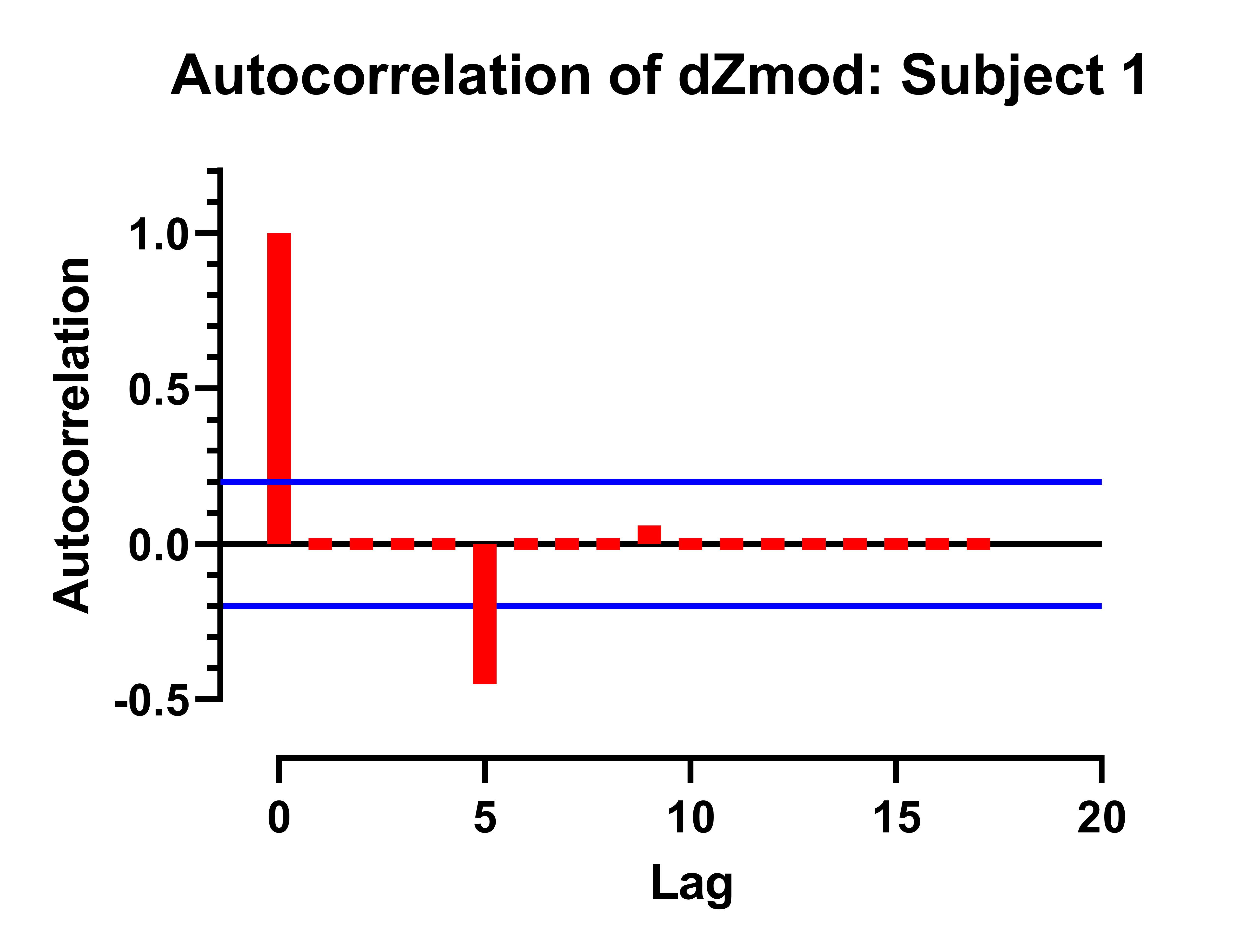} }\label{acfdZ1}}
    \quad
    \subfloat[]{{\includegraphics[width=0.42\textwidth]{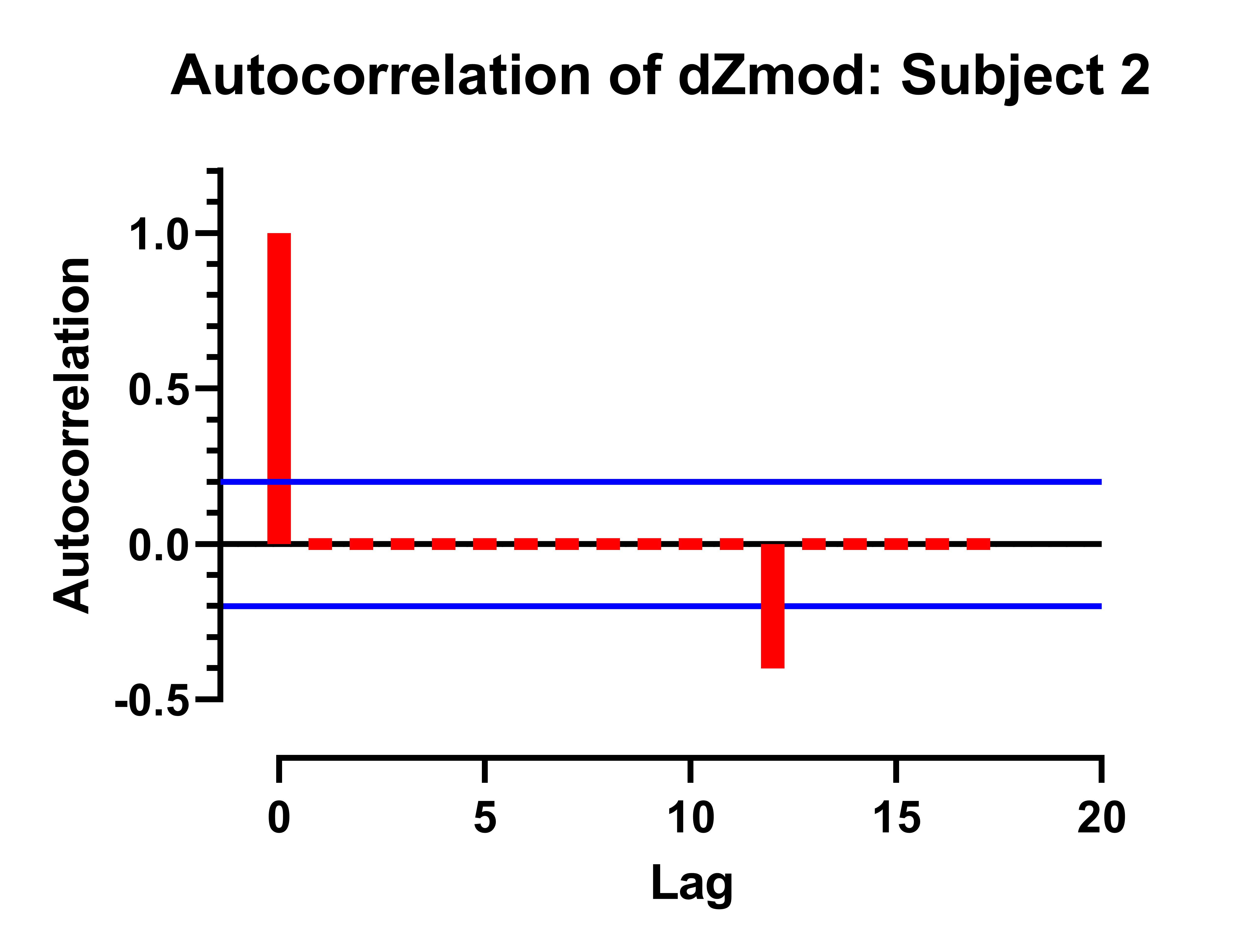} }\label{acfdZ2}}
    \quad
    
    \subfloat[]{{\includegraphics[width=0.42\textwidth]{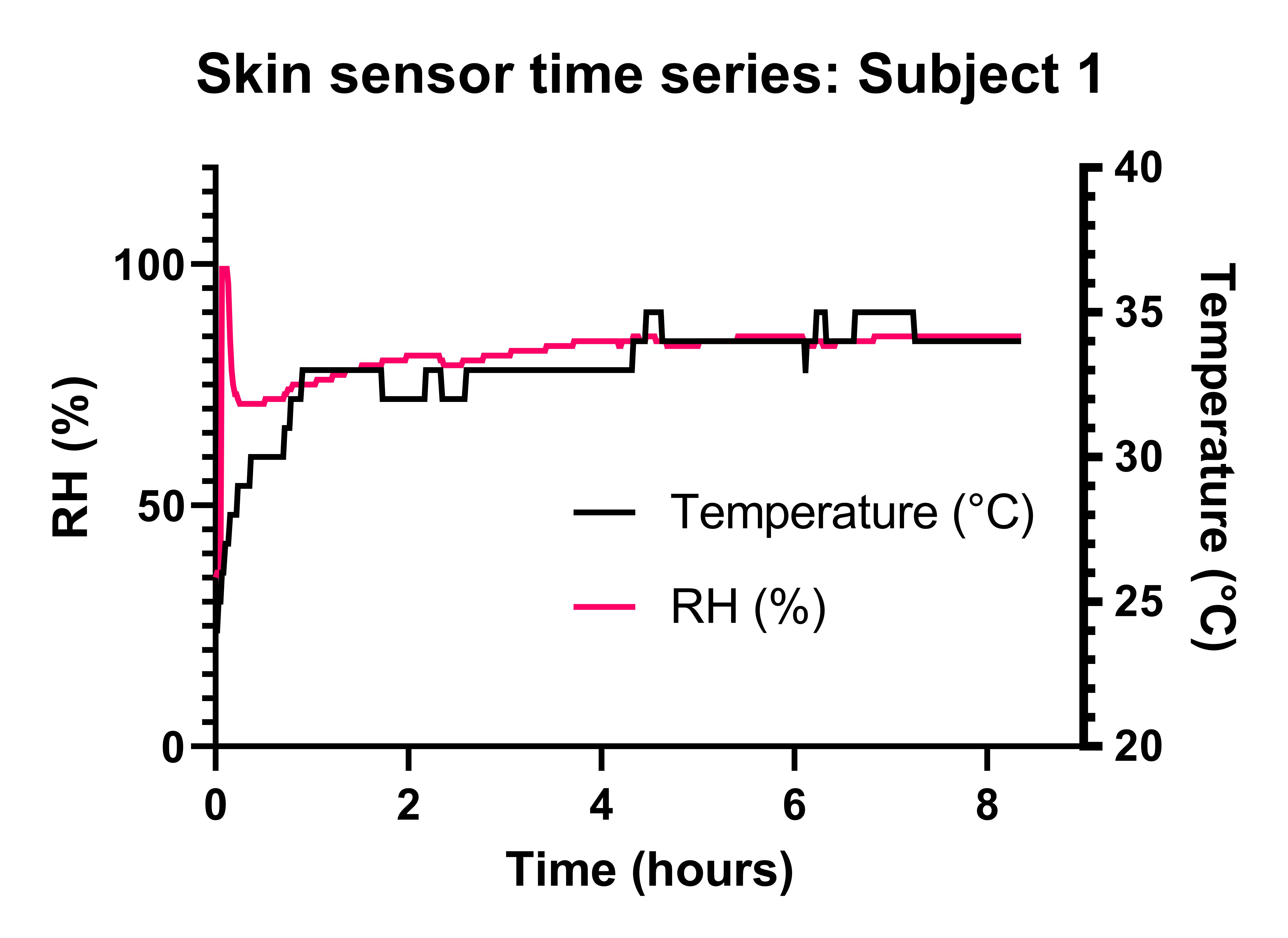} }\label{skin1}}
    \quad3
    \subfloat[]{{\includegraphics[width=0.42\textwidth]{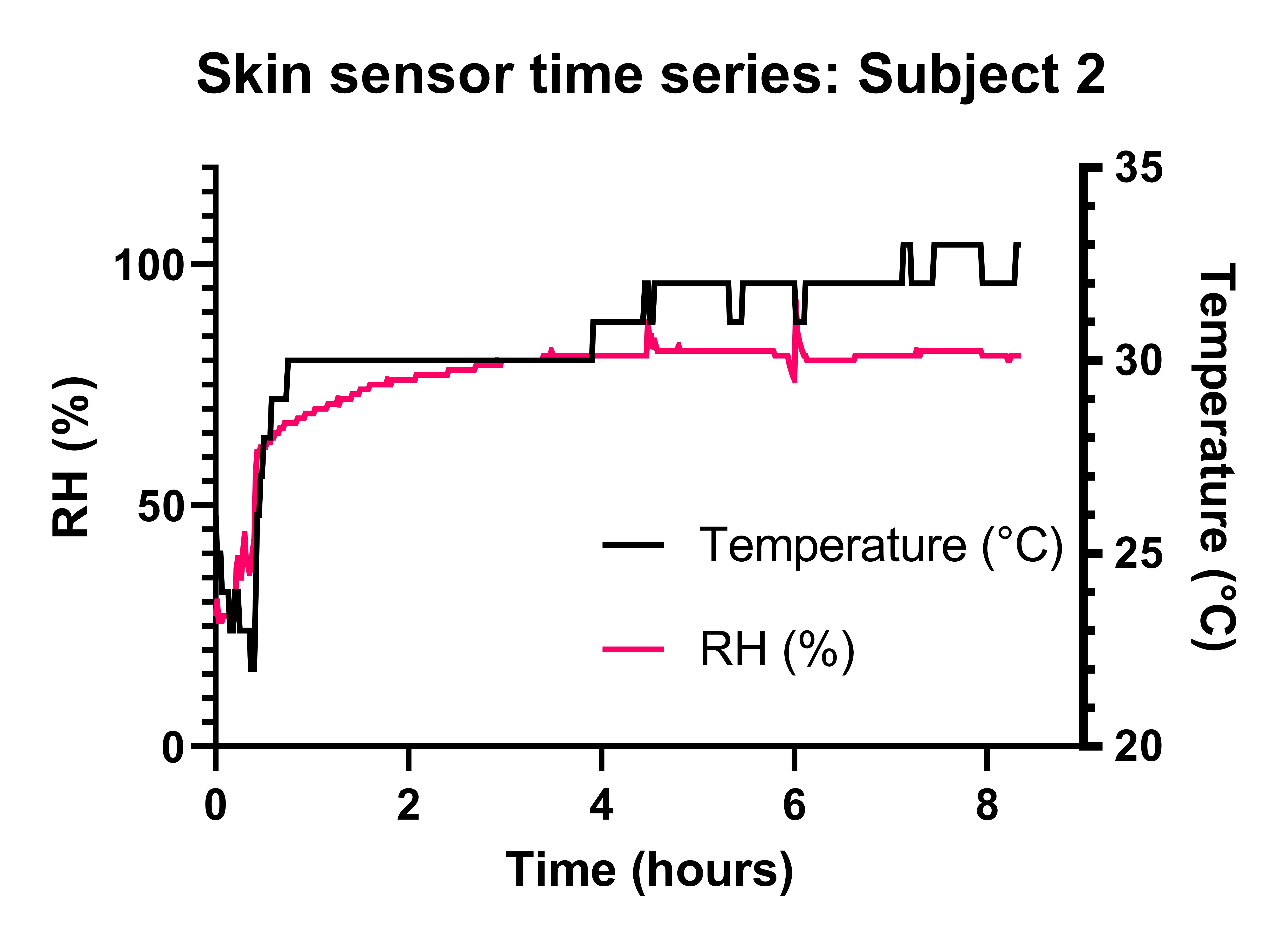} }\label{skin2}}
    
    \caption{~\ref{acfZ1} Autocorrelation plot for Zmod for subject 1.
    ~\ref{acfZ2} Autocorrelation plot for Zmod for subject 2.
    ~\ref{acfdZ1} Autocorrelation plot for dZmod for subject 1.
    ~\ref{acfdZ2} Autocorrelation plot for dZmod for subject 2.
    ~\ref{skin1} Skin temperature and relative humidity for subject 1.
    ~\ref{skin2} Skin temperature and relative humidity for subject 2.
    }\label{ARIMAsub}
\end{figure}

ARIMA(p,d,q) is one of the most commonly used curve fit functions used for time-series analysis, where $p$ is the order of auto-regressive (AR) terms, $d$ is the number of integrating terms and $q$ is the number of moving -average (MA) terms. For the above data set, $p$ was determined using the number of significant peaks in the auto-correlation plot of the primary predictors impedance magnitude (Zmod) and first differential of impedance magnitude (dZmod), which shows a change in sensor response with change in glucose concentration as per dose calibration. Figs.~\ref{acfZ1},\ref{acfdZ2} show 2 extreme cases among the 20 data set, wherein the auto-correlation plot shows 5 initial lag peaks for subject 1 and 10 initial lag peaks for subject 2, which exceed the $3\sigma$ confidence interval. Hence, the maximum peaks defined $p=10$. The ARIMA(10,0,3) regression curve fit used for the above data sets is given in Eqn.~\ref{ARIMAeqn1} and Eqn.~\ref{ARIMAeqn2}, where $y$ is the output, $X$ are the input values for the predictors, $\beta$ are the coefficients obtained from regression ARIMA fit, $\mu_t$ is the variance, $L$ is the lag operator, $\theta$ are the moving average coefficients and $\epsilon_t$ is Gaussian white noise. 

\begin{figure}
\centering
	\subfloat[]{{\includegraphics[width=0.42\textwidth]{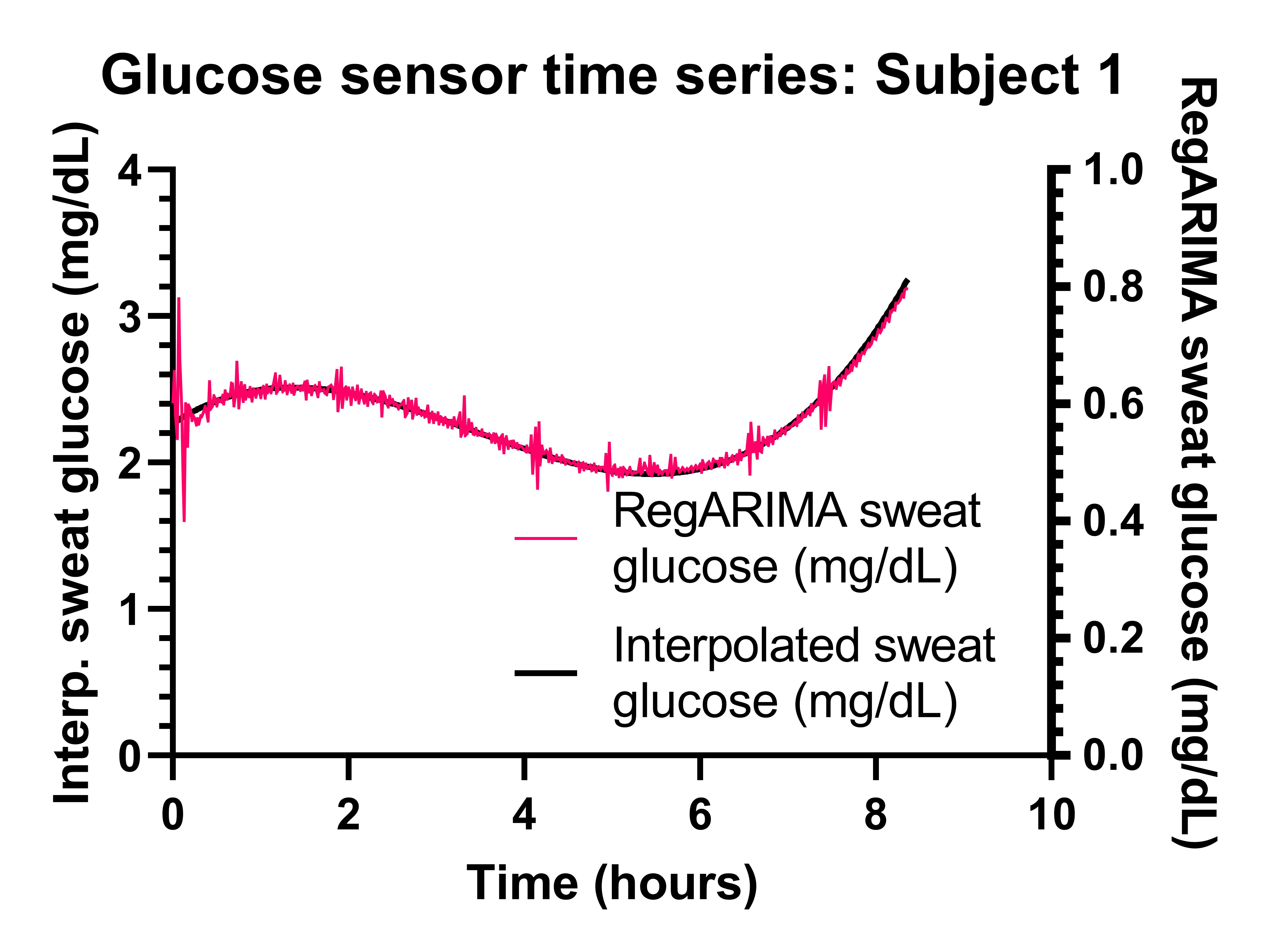} }\label{g1}}
    \quad
    \subfloat[]{{\includegraphics[width=0.42\textwidth]{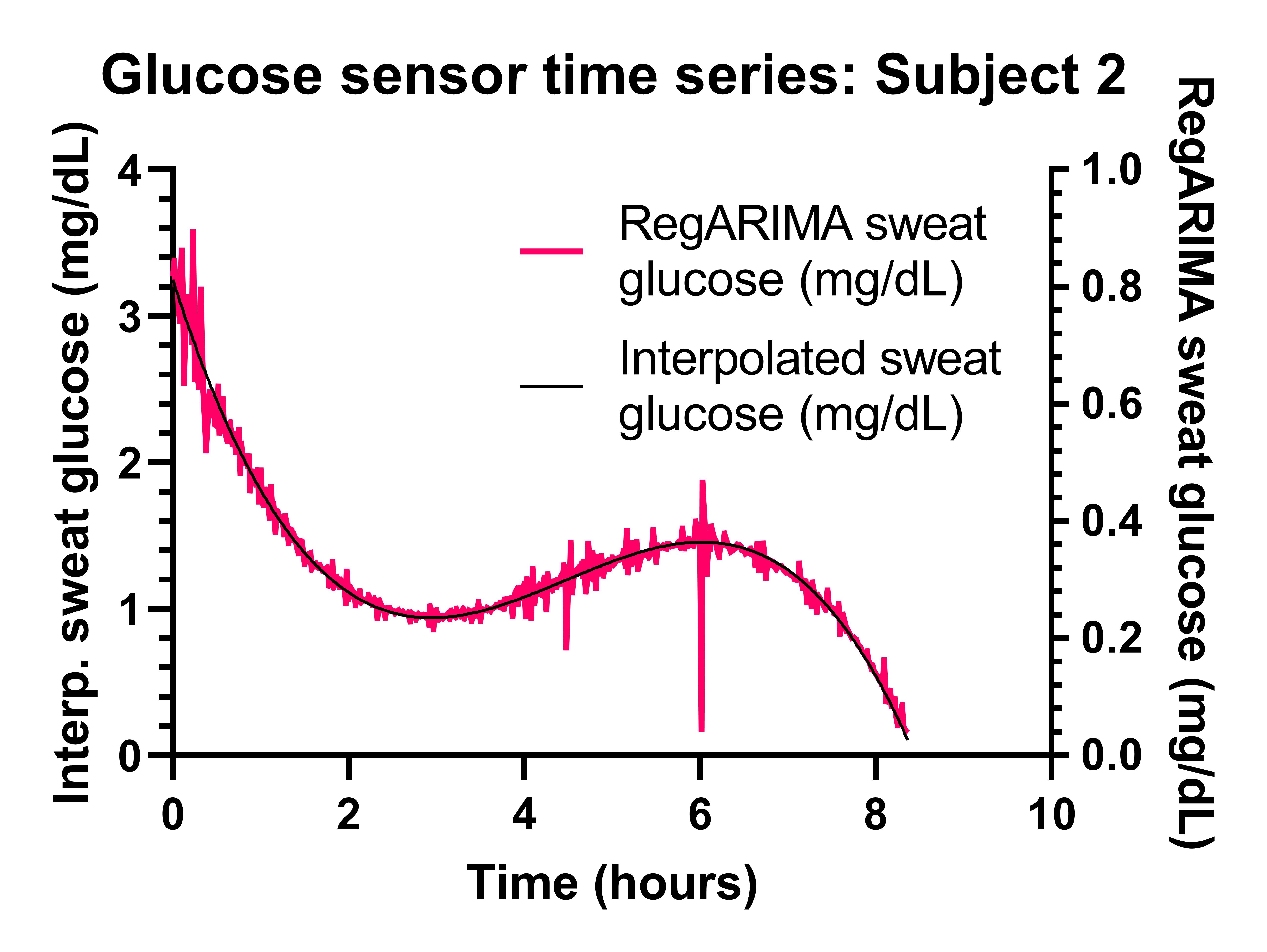} }\label{g2}}
    \quad
    
    \subfloat[]{{\includegraphics[width=0.42\textwidth]{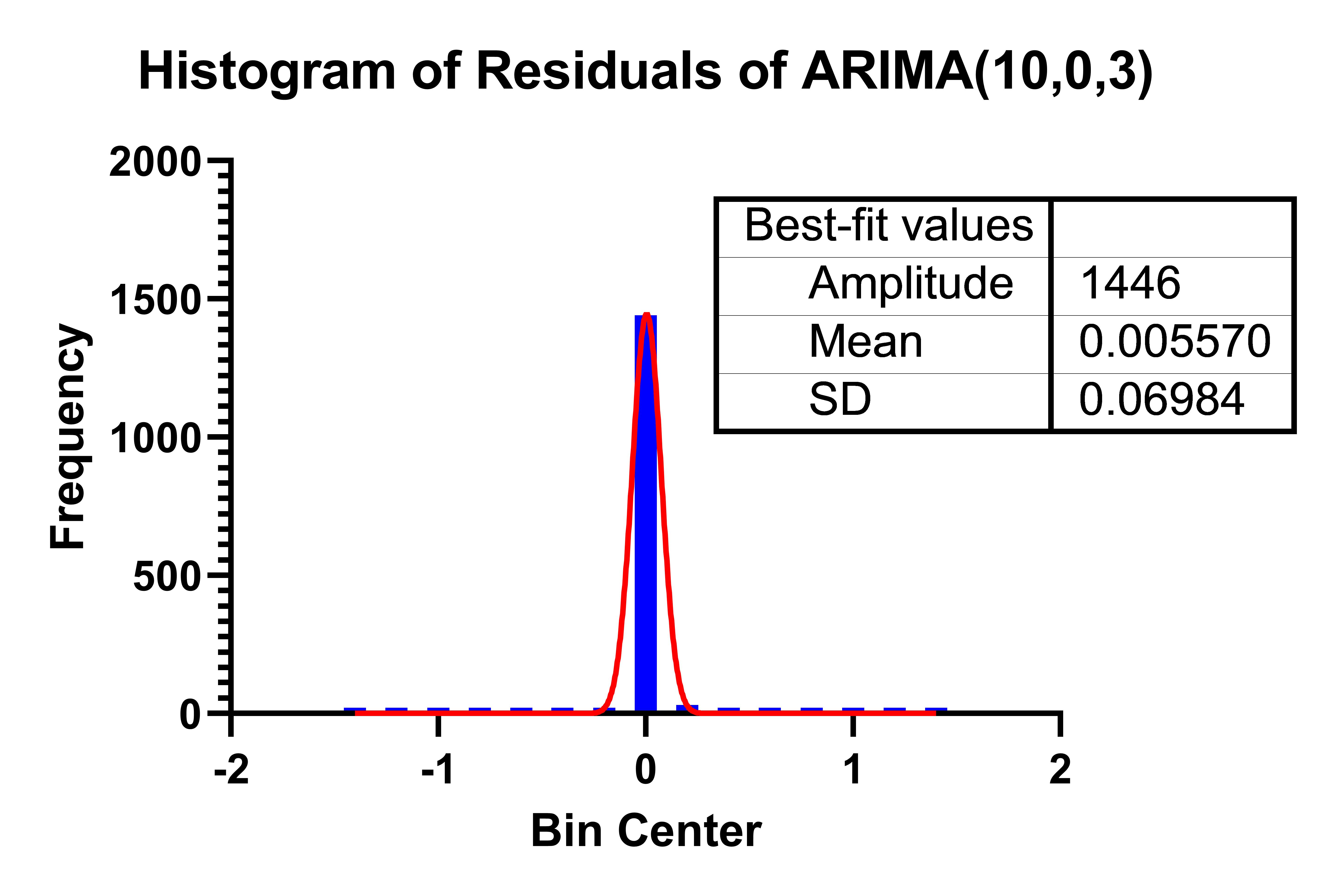} }\label{hist}}
    
    \caption{~\ref{g1} Comparison of interpolated vs. RegARIMA sweat glucose concentrations for subject 1.
    ~\ref{g2} Comparison of interpolated vs. RegARIMA sweat glucose concentrations for subject 2.
    ~\ref{hist} Histogram of residuals for ARIMA(10,0,3) model. 
    }\label{ARIMAout}
\end{figure}

The coefficients of the ARIMA(10,0,3) curve fit are shown in Table~\ref{tab:arima}. The p-value for a given predictor defines its significance in predicting the output time series, which is the interpolated sweat glucose value. If $p<0.05$, then the predictor is assumed to be significant as it rejects the null hypothesis, else the predictor can be assumed to be non-significant, hence can be removed from the curve fit expression. In this case, predictors $AR4$ can be safely assumed to be not contributing to the prediction. The AIC of this curve fit was $-3131.438$, ensuring good curve fit and low prediction error. This is corroborated by the ARIMA outputs plotted against the interpolated glucose values in Figs.~\ref{g1},~\ref{g2}. The error histogram is shown in Fig.~\ref{hist}, demonstrating a low predicted sweat glucose error with respect to the interpolated sweat glucose measurements.

\begin{table}[htb]
\centering
\begin{tabular}{|p{1in}|p{1in}|p{1in}|p{1in}|p{1in}|} \hline
\textbf{Predictor} & \textbf{Value} & \textbf{Std. Error} & \textbf{t-stat} & \textbf{p-value} \\ \hline
Intercept & 2.2236 & 0.0011792 & 1885.6396 & 0\\\hline AR{1} & 1.1696 & 0.016376 & 71.4217 & 0\\\hline AR{2} & -0.79248 & 0.015316 & -51.7426 & 0\\\hline AR{3} & 0.71839 & 0.044644 & 16.0916 & 2.9197e-58\\\hline AR{4} & 0.10234 & 0.044076 & 2.322 & 0.020235\\\hline AR{5} & -0.22962 & 0.032213 & -7.1282 & 1.0166e-12\\\hline AR{6} & 0.1 & 0.034417 & 2.9055 & 0.0036662\\\hline AR{7} & -0.20436 & 0.034488 & -5.9257 & 3.109e-09\\\hline AR{8} & 0.11278 & 0.03253 & 3.4668 & 0.00052668\\\hline AR{9} & -0.075336 & 0.032725 & -2.3021 & 0.021328\\\hline AR{10} & 0.085203 & 0.026248 & 3.246 & 0.0011702\\\hline MA{1} & -0.60583 & 0.017528 & -34.5629 & 9.1092e-262\\\hline MA{2} & 0.7133 & 0.024382 & 29.255 & 3.8771e-188\\\hline MA{3} & -0.40103 & 0.050977 & -7.8668 & 3.6377e-15\\\hline Beta(Zimag) & -0.00025308 & 4.887e-06 & -51.7866 & 0\\\hline Beta(Zmod) & 0.00054978 & 1.1027e-05 & 49.8589 & 0\\\hline Beta(Zphase) & -0.023363 & 0.0014868 & -15.7133 & 1.2261e-55\\\hline Beta(Zreal) & -0.00047466 & 9.8819e-06 & -48.033 & 0\\\hline Beta(dZimag) & -5.2582e-06 & 6.8561e-07 & -7.6694 & 1.7285e-14\\\hline Beta(dZmod) & -7.0023e-06 & 1.5741e-07 & -44.4853 & 0\\\hline Beta(dZphase) & -0.0045203 & 0.0022546 & -2.0049 & 0.044974\\\hline Beta(dZreal) & 1.1458e-05 & 6.5176e-07 & 17.5794 & 3.5439e-69\\\hline Beta(rh) & -0.022069 & 0.00070336 & -31.3763 & 4.2615e-216\\\hline Variance & 0.0075866 & 0.00025797 & 29.4091 & 4.2045e-190\\\hline 
\end{tabular}
\caption{\label{tab:arima}Values of ARIMA regression coefficients along with standard error, t-statistic and p-value as obtained from fitting on test data sets.}
\end{table}

\begin{table}[H]
\hyphenation{response}
\hyphenation{generator}
\centering
\begin{tabular}{|p{1.5in}|p{1.5in}|p{1.5in}|p{1.5in}|}
\hline
\textbf{Specification} & \textbf{Jalal et. al.~\cite{bhansali}} & \textbf{Lee et. al.~\cite{lee2016graphene} } & \textbf{This work}\\
\hline
Sensing mechanism &	chronoamperometry & chronoamperometry with potential pre-stabilization & EIS\\
\hline
Front end & LMP91000 + reference generator + DAC + passive 2nd order filter & MAX4617 (analog mux) + MAX7405 (8th order filter) + AFE4490 (amperometric front end) & ADuCM350\\
\hline
Sensor + electronic response time & 200 s (OCP) + 200 s (chronoamp) & 60 s (pre stabilization) + 180 s & 180 s + 13 ms\\
\hline
Processor&	TI MSP430F5229 & ATMEGA328 & ADuCM350\\
\hline
Sensing algorithm &	hysteresis (hard thresholds) on OCP + chronoamperometry	& temperature and PPS based normalization (guarantees accuracy ) & auto-regression\\
\hline
Battery size &	1000 mAh &	N/A & 300 mAh\\
\hline
Peripherals	& Bluetooth, BMS, EEPROM & Bluetooth, BMS, SpO2, PPG, temperature & BLE, BMS, temperature, RH\\
\hline
\end{tabular}
\caption{\label{tab:compare}Electrical comparison of specifications of system level designs reported in literature.}
\end{table}

\section*{Discussion}
Various electrochemical techniques such as cyclic voltammetry, chronoamperometry and impedance spectroscopy techniques exist for the characterization of electrochemical cells and have been successfully employed for biosensing applications. However, there has been less interest in design impedance spectroscopy device due to increased complexity required for phase as well as magnitude detection on the hardware aspect. Moreover, added hardware increases the size and power consumption on a system level scale for most designs. Thus, the first wearable, electrochemical impedance spectroscopy system is proposed for glucose detection in sweat with a battery lifetime of 215 hours. Moreover, a small form factor of the device is demonstrated by implementing a monolithic SoC with an EIS dedicated front-end with four channel sensing. Table~\ref{tab:compare} compares the electrical specifications and system level mechanism in general to highlight the advantages of using an impedance based spectroscopy design over an amperometric design.

The proposed biosensor is capable of sensing glucose concentrations in sweat from 10 mg/dL to 200 mg/dL in various synthetic sweat buffers of various pH values. Moreover, the stability of the sensor response is established not only from a conventional statistical approach, but also using sample-to-sample variation observations and demonstrates a 20\% variability similar to safe limits on a Clarke Error Grid. In addition, for the first time, a semi-quantitative method of detecting rate of change of glucose is proposed for aiding real-time diagnostic efforts using a slope feature extraction algorithm on the impedance response of the sensor. This algorithm also demonstrates a rate-of-change occurrence probability-based reduction in time complexity, leading to low power operation on a wearable platform.

From a system level perspective, this work is a first attempt to dive into the 
specifics of interdisciplinary system integration for biosensing applications. A simplistic noise model for the biosensor is demonstrated along with noise measurements to correlate with change of noise behavior in time based on conventional electrical circuit noise analyses. The analysis is tested against noise measurements made on a high-speed potentiostat and mathematically verified. In addition, the accuracy of the measurement device is identified and coupled to a source of error which corroborates the MARD based analysis of sample-to-sample variability. Finally, a mismatch expression is proposed to demonstrate low mismatch due to coupling of the biosensor and the measurement device with a mismatch coefficient of -0.875\%.

\section*{Methods}

\subsection*{Sensor fabrication}
Fig.~\ref{molecular} shows the cross-sectional view of the chemi-impedance biosensor and the formation of the immunoassay on the functionalized surface. The biosensor substrate is a 100 $\mu$m thick,flexible nanoporous polyamide substrate (GE Lifesciences, NJ). Gold electrodes were patterned using shadow mask and deposited using e-beam cryo-evaporator. ZnO thin films were sputtered onto patterned in the area between the two gold electrodes to get maximum overlap using an AJA Orion RF magnetron with a 99.999\% ZnO target (Kurt J. Lesker) at room temperature~\cite{rm_glucose}. 

\subsection*{Materials for implementing glucose assay}
Poly-amide substrates with 0.2 $\mu$m pore size were obtained
from GE Healthcare Life Sciences (Piscataway, NJ, USA). The linker
molecule dithiobis [succinimidyl propionate] (DSP) and its solvent
dimethyl sulfoxide (DMSO) were ordered from Thermo Fisher
Scientific Inc. (Waltham, MA, USA). The monoclonal glucose oxidase
antibody was obtained from Thermo Fisher Scientific Inc.
(Waltham, MA, USA). Glucose oxidase from Aspergillus niger and D(+)-glucose was obtained from Sigma-Aldrich (St. Louis, MO, USA). The glucose antibody was diluted in 1X phosphate buffered saline (PBS, Thermo Fisher Scientific Inc., Waltham, MA, USA). Synthetic sweat was prepared as per the recipe stated in M.T. Mathew et al. The pH range is varied by varying the concentration of the components. Human sweat was purchased from Lee Biosolutions Inc. (St. Louis, MO, USA), where it was collected from single human donor with pH $\approx$ 4–5. No preservatives have been added to this product and it was stored unfiltered at below - 20 \textdegree C~\cite{rm_glucose}.

\subsection*{Wearable EIS device disassembly}
The wearable EIS device used for experiments in this work is shown in Fig.~\ref{mbx}. The design is primarily divided into two parts: a mother board and a daughter board. The motherboard houses the integrated circuits required for measurement and communication. The daughter board houses four flexible biosensors using slide-in connectors. The daughter board connects to the motherboard using a slide-in connector as well. The motherboard is further divided into four subsections. The first subsection is the ADuCM350 system-on-chip (SoC) (Analog Devices Inc.) which is the heart of the wearable platform. It is equipped with an impedance analyzer block which gives the real and imaginary part of an unknown impedance using a ratiometric discrete Fourier transform (DFT) accelerator. The second subsection is the nRF8001 (Nordic Semiconductor ASA) Bluetooth Low Energy (BLE) chip-set which communicates with a mobile application. The third subsection is the temperature and RH sensing which was implemented using HDC1080 (Texas Instruments Inc.). The fourth subsection is the power management system implemented using BQ24040 (Texas Instruments Inc.). This subsections supplies power to the other subsection either using a lithium battery or USB and manages the charge-discharge cycle of the lithium battery.

\subsection*{Glucose sensor experimental protocol}
Gold surface was functionalized with 3 $\mu$L of 10 mmol DSP linker after incubation for 2 hours. The PBS wash was carried out followed by 15 min incubation of 3 $\mu$L of 10 $\mu$g/mL glucose oxidase antibody. Thereafter, 3 $\mu$L of 1 mg/mL glucose oxidase enzyme obtained from Aspergillus Niger is dispensed. Serial dilutions of glucose in synthetic sweat were prepared to range from 5 mg/dL to 200 mg/dL, which were dispensed after antibody incubation. Each glucose concentration was incubated for 5 minutes on the sensor surface prior to measurement. Sensor calibration response was obtained in human sweat with increased glucose concentration doses from 5 mg/dL to 200 mg/dL with n=4 replicates of chips. Two test sample chips with 5 mg/mL and 200 mg/mL glucose concentration doses on each were also measured. The average readings were then compared with the values obtained from calibration dose response in human sweat. These EIS measurements were performed by recording current flow using ADuCM350 after applying 10 mV AC voltage with a frequency of 100 Hz.


\subsection*{Human subject enrollment}
Human subjects for sweat sample collection and on-body continuous measurement in compliance with the protocol approved by Institutional Review Board (IRB) at the University of Texas at Dallas (IRB number 19-146). A written and informed consent was obtained from all participants of this study. A study was performed on 20 subjects to collect sedentary sweat glucose, skin temperature and RH over a period of 8 hours in compliance with the (IRB number 19-146) approved at the University of Texas at Dallas. All subjects belonged to the age range of 18-40 years with no prior diagnosis of diabetes mellitus. Each subject was made to follow the following schedule: The subject will come in at T0 and has breakfast at T0.5 and blood glucose measures are taken at T0 and T1, where n is the number of hours for a time instant Tn. The subject will have lunch at T3, and the blood glucose will be observed at T4. A final blood glucose measure will be taken at T8 to capture the fall in blood glucose over the duration of experiment. The output of the proposed device and prediction system was validated against the interpolated sweat glucose sample collected separately using a MacroDuct\textregistered sweat collector at above mentioned time points. 

\subsection*{On-body measurement of human subjects}
A written and informed consent was taken from all participants prior to sample collection. All enrolled subjects wore the devices during the duration of the testing and continuous measurements were recorded in compliance with the protocol approved by the IRB committee. The wearable sweat sensor device was placed on the antebrachial region (lower forearm) of the subject. The human subjects were neither allowed to exert themselves physically nor their skin was excited using a sweat induction method such as iontophoresis. This ensured a full passive generation of sweat to be sampled by AWARE. Continuous on-body measurements sampled every minute for the duration of the testing and depending on the availability of the subject. The concentration profile of the biomarker was reported over the entire time period of recording.

\section*{Acknowledgements}
The authors acknowledge David Kinnamon and Paul Rice for technical support on sensor fabrication and mechanical design.

\section*{Author contributions statement}

S.P. and S.M. conceived the experiments, K.C.L. and J.B. performed glucose sensor dose calibration, M.P. conducted the human subject data collection and validation, D.S. conducted the electrical characterization test experiments, curated all human subject data and created the AR model, all authors analyzed the results and reviewed the manuscript. 

\section*{Additional information} 



Drs. Shalini Prasad and Sriram Muthukumar have a significant interest in Enlisense LLC, a company that may have a commercial interest in the results of this research and technology. The potential individual conflict of interest has been reviewed and managed by The University of Texas at Dallas, and played no role in the study design; in the collection, analysis, and interpretation of data; in the writing of the report, or in the decision to send the report for publication. Authors D.S, M.P., K.C.L. and J.B. declare no competing financial interests. 

\bibliography{main}

\begin{thebibliography}{10}
\expandafter\ifx\csname url\endcsname\relax
  \def\url#1{\texttt{#1}}\fi
\expandafter\ifx\csname urlprefix\endcsname\relax\def\urlprefix{URL }\fi
\expandafter\ifx\csname doiprefix\endcsname\relax\def\doiprefix{DOI }\fi
\providecommand{\bibinfo}[2]{#2}
\providecommand{\eprint}[2][]{\url{#2}}

\bibitem{wang2001glucose}
\bibinfo{author}{Wang, J.}
\newblock \bibinfo{journal}{\bibinfo{title}{Glucose biosensors: 40 years of
  advances and challenges}}.
\newblock {\emph{\JournalTitle{Electroanalysis}}}
  \textbf{\bibinfo{volume}{13}}, \bibinfo{pages}{983} (\bibinfo{year}{2001}).

\bibitem{lee2016graphene}
\bibinfo{author}{Lee, H.} \emph{et~al.}
\newblock \bibinfo{journal}{\bibinfo{title}{A graphene-based electrochemical
  device with thermoresponsive microneedles for diabetes monitoring and
  therapy}}.
\newblock {\emph{\JournalTitle{Nature nanotechnology}}}
  \textbf{\bibinfo{volume}{11}}, \bibinfo{pages}{566} (\bibinfo{year}{2016}).

\bibitem{bhansali}
\bibinfo{author}{Jalal, A.~H.}, \bibinfo{author}{Umasankar, Y.},
  \bibinfo{author}{Gonzalez, P.~J.}, \bibinfo{author}{Alfonso, A.} \&
  \bibinfo{author}{Bhansali, S.}
\newblock \bibinfo{journal}{\bibinfo{title}{Multimodal technique to eliminate
  humidity interference for specific detection of ethanol}}.
\newblock {\emph{\JournalTitle{Biosensors and Bioelectronics}}}
  \textbf{\bibinfo{volume}{87}}, \bibinfo{pages}{522--530}
  (\bibinfo{year}{2017}).

\bibitem{kimAlcohol}
\bibinfo{author}{Kim, J.} \emph{et~al.}
\newblock \bibinfo{journal}{\bibinfo{title}{Noninvasive alcohol monitoring
  using a wearable tattoo-based iontophoretic-biosensing system}}.
\newblock {\emph{\JournalTitle{ACS Sensors}}} \textbf{\bibinfo{volume}{1}},
  \bibinfo{pages}{1011--1019} (\bibinfo{year}{2016}).

\bibitem{gaoMUX}
\bibinfo{author}{Gao, W.} \emph{et~al.}
\newblock \bibinfo{journal}{\bibinfo{title}{Fully integrated wearable sensor
  arrays for multiplexed in situ perspiration analysis}}.
\newblock {\emph{\JournalTitle{Nature}}} \textbf{\bibinfo{volume}{529}},
  \bibinfo{pages}{509} (\bibinfo{year}{2016}).

\bibitem{davidSWRalcohol}
\bibinfo{author}{Kinnamon, D.}, \bibinfo{author}{Selvam, A.~P.},
  \bibinfo{author}{Prasad, S.} \& \bibinfo{author}{Muthukumar, S.}
\newblock \bibinfo{title}{Electronic bracelet for monitoring of alcohol
  lifestyle}.
\newblock In \emph{\bibinfo{booktitle}{2016 IEEE SENSORS}},
  \bibinfo{pages}{1--3} (\bibinfo{year}{2016}).
\newblock \doiprefix 10.1109/ICSENS.2016.7808598.

\bibitem{cgmstdBattelino2019clinical}
\bibinfo{author}{Battelino, T.} \emph{et~al.}
\newblock \bibinfo{journal}{\bibinfo{title}{Clinical targets for continuous
  glucose monitoring data interpretation: recommendations from the
  international consensus on time in range}}.
\newblock {\emph{\JournalTitle{Diabetes care}}} \textbf{\bibinfo{volume}{42}},
  \bibinfo{pages}{1593--1603} (\bibinfo{year}{2019}).

\bibitem{bardFaukner}
\bibinfo{author}{Bard, A.~J.}, \bibinfo{author}{Faulkner, L.~R.},
  \bibinfo{author}{Leddy, J.} \& \bibinfo{author}{Zoski, C.~G.}
\newblock \emph{\bibinfo{title}{Electrochemical methods: fundamentals and
  applications}}, vol.~\bibinfo{volume}{2} (\bibinfo{publisher}{wiley New
  York}, \bibinfo{year}{1980}).

\bibitem{rm_glucose}
\bibinfo{author}{Munje, R.~D.}, \bibinfo{author}{Muthukumar, S.} \&
  \bibinfo{author}{Prasad, S.}
\newblock \bibinfo{journal}{\bibinfo{title}{Lancet-free and label-free
  diagnostics of glucose in sweat using zinc oxide based flexible
  bioelectronics}}.
\newblock {\emph{\JournalTitle{Sensors and Actuators B: Chemical}}}
  \textbf{\bibinfo{volume}{238}}, \bibinfo{pages}{482 -- 490}
  (\bibinfo{year}{2017}).
\newblock \doiprefix https://doi.org/10.1016/j.snb.2016.07.088.

\bibitem{SankhalaCortisol}
\bibinfo{author}{Sankhala, D.}, \bibinfo{author}{Muthukumar, S.} \&
  \bibinfo{author}{Prasad, S.}
\newblock \bibinfo{journal}{\bibinfo{title}{A four-channel electrical impedance
  spectroscopy module for cortisol biosensing in sweat-based wearable
  applications}}.
\newblock {\emph{\JournalTitle{SLAS TECHNOLOGY: Translating Life Sciences
  Innovation}}} \textbf{\bibinfo{volume}{0}}, \bibinfo{pages}{2472630318759257}
  (\bibinfo{year}{0}).
\newblock \doiprefix 10.1177/2472630318759257.
\newblock \bibinfo{note}{PMID: 29447045}.

\bibitem{Hooge}
\bibinfo{author}{Hooge, F.~N.}
\newblock \bibinfo{journal}{\bibinfo{title}{1/f noise sources}}.
\newblock {\emph{\JournalTitle{IEEE Transactions on Electron Devices}}}
  \textbf{\bibinfo{volume}{41}}, \bibinfo{pages}{1926--1935}
  (\bibinfo{year}{1994}).
\newblock \doiprefix 10.1109/16.333808.

\bibitem{thermalNoise}
\bibinfo{author}{Johnson, J.~B.}
\newblock \bibinfo{journal}{\bibinfo{title}{Thermal agitation of electricity in
  conductors}}.
\newblock {\emph{\JournalTitle{Phys. Rev.}}} \textbf{\bibinfo{volume}{32}},
  \bibinfo{pages}{97--109} (\bibinfo{year}{1928}).
\newblock \doiprefix 10.1103/PhysRev.32.97.

\bibitem{razavi}
\bibinfo{author}{Razavi, B.}
\newblock \emph{\bibinfo{title}{Design of Analog CMOS Integrated Circuits}}
  (\bibinfo{publisher}{McGraw-Hill, Inc.}, \bibinfo{address}{New York, NY,
  USA}, \bibinfo{year}{2001}), \bibinfo{edition}{1} edn.

\bibitem{hassibi_noise}
\bibinfo{author}{Hassibi, A.}, \bibinfo{author}{Navid, R.},
  \bibinfo{author}{Dutton, R.~W.} \& \bibinfo{author}{Lee, T.~H.}
\newblock \bibinfo{journal}{\bibinfo{title}{Comprehensive study of noise
  processes in electrode electrolyte interfaces}}.
\newblock {\emph{\JournalTitle{Journal of Applied Physics}}}
  \textbf{\bibinfo{volume}{96}}, \bibinfo{pages}{1074--1082}
  (\bibinfo{year}{2004}).
\newblock \doiprefix 10.1063/1.1755429.

\bibitem{znonoiseBarhoumi2014correlations}
\bibinfo{author}{Barhoumi, A.} \emph{et~al.}
\newblock \bibinfo{journal}{\bibinfo{title}{Correlations between 1/f noise and
  thermal treatment of al-doped zno thin films deposited by direct current
  sputtering}}.
\newblock {\emph{\JournalTitle{Journal of Applied Physics}}}
  \textbf{\bibinfo{volume}{115}}, \bibinfo{pages}{204502}
  (\bibinfo{year}{2014}).

\bibitem{NoujaimMARD}
\bibinfo{author}{Noujaim, S.~E.}, \bibinfo{author}{Horwitz, D.},
  \bibinfo{author}{Sharma, M.} \& \bibinfo{author}{Marhoul, J.}
\newblock \bibinfo{journal}{\bibinfo{title}{Accuracy requirements for a
  hypoglycemia detector: An analytical model to evaluate the effects of bias,
  precision, and rate of glucose change}}.
\newblock {\emph{\JournalTitle{Journal of Diabetes Science and Technology}}}
  \textbf{\bibinfo{volume}{1}}, \bibinfo{pages}{652--668}
  (\bibinfo{year}{2007}).
\newblock \doiprefix 10.1177/193229680700100509.
\newblock \bibinfo{note}{PMID: 19885133}.

\bibitem{WaldenADC}
\bibinfo{author}{Walden, R.~H.}
\newblock \bibinfo{journal}{\bibinfo{title}{Analog-to-digital converter survey
  and analysis}}.
\newblock {\emph{\JournalTitle{IEEE Journal on Selected Areas in
  Communications}}} \textbf{\bibinfo{volume}{17}}, \bibinfo{pages}{539--550}
  (\bibinfo{year}{1999}).
\newblock \doiprefix 10.1109/49.761034.

\bibitem{leal2010real}
\bibinfo{author}{Leal, Y.} \emph{et~al.}
\newblock \bibinfo{journal}{\bibinfo{title}{Real-time glucose estimation
  algorithm for continuous glucose monitoring using autoregressive models}}.
\newblock {\emph{\JournalTitle{Journal of diabetes science and technology}}}
  \textbf{\bibinfo{volume}{4}}, \bibinfo{pages}{391--403}
  (\bibinfo{year}{2010}).

\bibitem{reifman2007predictive}
\bibinfo{author}{Reifman, J.}, \bibinfo{author}{Rajaraman, S.},
  \bibinfo{author}{Gribok, A.} \& \bibinfo{author}{Ward, W.~K.}
\newblock \bibinfo{journal}{\bibinfo{title}{Predictive monitoring for improved
  management of glucose levels}}.
\newblock {\emph{\JournalTitle{Journal of diabetes science and technology}}}
  \textbf{\bibinfo{volume}{1}}, \bibinfo{pages}{478--486}
  (\bibinfo{year}{2007}).

\bibitem{dash2020neurovad}
\bibinfo{author}{Dash, D.}, \bibinfo{author}{Ferrari, P.},
  \bibinfo{author}{Dutta, S.} \& \bibinfo{author}{Wang, J.}
\newblock \bibinfo{journal}{\bibinfo{title}{Neurovad: Real-time voice activity
  detection from non-invasive neuromagnetic signals}}.
\newblock {\emph{\JournalTitle{Sensors}}} \textbf{\bibinfo{volume}{20}},
  \bibinfo{pages}{2248} (\bibinfo{year}{2020}).

\bibitem{mhaskar2017deep}
\bibinfo{author}{Mhaskar, H.~N.}, \bibinfo{author}{Pereverzyev, S.~V.} \&
  \bibinfo{author}{van~der Walt, M.~D.}
\newblock \bibinfo{journal}{\bibinfo{title}{A deep learning approach to
  diabetic blood glucose prediction}}.
\newblock {\emph{\JournalTitle{Frontiers in Applied Mathematics and
  Statistics}}} \textbf{\bibinfo{volume}{3}}, \bibinfo{pages}{14}
  (\bibinfo{year}{2017}).

\bibitem{naumova2012meta}
\bibinfo{author}{Naumova, V.}, \bibinfo{author}{Pereverzyev, S.~V.} \&
  \bibinfo{author}{Sivananthan, S.}
\newblock \bibinfo{journal}{\bibinfo{title}{A meta-learning approach to the
  regularized learning—case study: Blood glucose prediction}}.
\newblock {\emph{\JournalTitle{Neural Networks}}}
  \textbf{\bibinfo{volume}{33}}, \bibinfo{pages}{181--193}
  (\bibinfo{year}{2012}).

\bibitem{sivananthan2011assessment}
\bibinfo{author}{Sivananthan, S.} \emph{et~al.}
\newblock \bibinfo{journal}{\bibinfo{title}{Assessment of blood glucose
  predictors: the prediction-error grid analysis}}.
\newblock {\emph{\JournalTitle{Diabetes technology \& therapeutics}}}
  \textbf{\bibinfo{volume}{13}}, \bibinfo{pages}{787--796}
  (\bibinfo{year}{2011}).

\bibitem{pappada2011neural}
\bibinfo{author}{Pappada, S.~M.} \emph{et~al.}
\newblock \bibinfo{journal}{\bibinfo{title}{Neural network-based real-time
  prediction of glucose in patients with insulin-dependent diabetes}}.
\newblock {\emph{\JournalTitle{Diabetes technology \& therapeutics}}}
  \textbf{\bibinfo{volume}{13}}, \bibinfo{pages}{135--141}
  (\bibinfo{year}{2011}).

\end{thebibliography}

\clearpage

\clearpage

\clearpage

\end{document}